\documentclass[%
 reprint,
superscriptaddress,
nofootinbib,
 amsmath,amssymb,
 aps,
prd,
]{revtex4-2}

\usepackage{graphicx}% Include figure files
\usepackage{dcolumn}% Align table columns on decimal point
\usepackage{bm}% bold math
\usepackage[english]{babel}
\usepackage{fancyhdr}
\usepackage{amsmath}
\usepackage{amssymb}
\usepackage{amsfonts}
\usepackage{psfrag}
\usepackage[applemac]{inputenc}
\usepackage[bf,footnotesize,justification=Justified,format=plain]{caption}
\usepackage[dvipsnames]{xcolor}
\usepackage[colorlinks=True, citecolor=blue, linkcolor=blue, urlcolor=blue,linktocpage]{hyperref}
\usepackage{amsthm}
\usepackage{gensymb}
\usepackage{subfig}
\usepackage{physics}
\usepackage{array}
\usepackage{tcolorbox, mathtools}
\usepackage{soul}
\usepackage{aas_macros}
\usepackage[normalem]{ulem}

\newcommand{\mpl}{M_\mathrm{P}}

\begin{document}

\title{Scalar emission from neutron star-black hole binaries in scalar-tensor theories with kinetic screening}% Force line breaks with \\

% \author{Ramiro Cayuso}
% \email{rcayuso@sissa.it}
%\author{Adrien Kuntz}
% \email{adrien.kuntz@sissa.it}
% \author{Enrico Barausse}
% \email{barausse@sissa.it}
%\affiliation{SISSA, Via Bonomea 265, 34136 Trieste, Italy and INFN Sezione di Trieste}
%\affiliation{IFPU - Institute for Fundamental Physics of the Universe, Via Beirut 2, 34014 Trieste, Italy}

\author{Ramiro Cayuso}
\email{rcayuso@sissa.it}

\author{Adrien Kuntz}
\email{adrien.kuntz@sissa.it}

\affiliation{SISSA, Via Bonomea 265, 34136 Trieste, Italy and INFN Sezione di Trieste}
\affiliation{IFPU - Institute for Fundamental Physics of the Universe, Via Beirut 2, 34014 Trieste, Italy}

\author{Miguel Bezares}  % New author in between
\email{miguel.bezaresfigueroa@nottingham.ac.uk}
\affiliation{Nottingham Centre of Gravity, Nottingham NG7 2RD, United Kingdom}  % New affiliation
\affiliation{School of Mathematical Sciences, University of Nottingham}  % New affiliation

\author{Enrico Barausse}
\email{barausse@sissa.it}

\affiliation{SISSA, Via Bonomea 265, 34136 Trieste, Italy and INFN Sezione di Trieste}
\affiliation{IFPU - Institute for Fundamental Physics of the Universe, Via Beirut 2, 34014 Trieste, Italy}

\date{\today}% It is always \today, today,
             %  but any date may be explicitly specified

\begin{abstract}
We explore scalar radiation from neutron star-black hole binaries in scalar-tensor theories with kinetic screening ($K$-essence). Using 3+1 numerical relativity simulations in the decoupling limit, we investigate scalar dipole and quadrupole radiation for different values of the strong coupling constant $\Lambda$. Our results show that kinetic screening effectively suppresses the scalar dipole radiation as $\Lambda$ decreases. 
This is validated by comparing to analytic predictions for the screening of dipole scalar emission, with which our numerical results show good agreement.
However, our numerical simulations show that the suppression of scalar quadrupole radiation is less efficient, even when the screening radius exceeds the wavelength of the emitted radiation. In fact, the dependence of the scalar quadrupole amplitude on $\Lambda$ flattens out for the smallest $\Lambda$ that
we can simulate, and the quadrupole amplitude 
is suppressed only by  a factor $\lesssim 3$ relative to the Fierz-Jordan-Brans-Dicke case.  Overall, our study shows that scalar quadrupole radiation from mixed binaries may be used to place constraints on $K$-essence theories with next-generation gravitational-wave detectors.
\end{abstract}

\maketitle

\section{Introduction}

Theories beyond General Relativity (GR) that introduce additional gravitational scalar degrees of freedom (scalar-tensor theories)~\cite{Fierz:1956zz,Jordan:1959eg,Brans:1961sx,Langlois:2015cwa, Crisostomi:2016czh,BenAchour:2016fzp}  have enjoyed renewed interest in recent years, particularly 
after the detection of gravitational waves (GWs). Indeed, these observations allow for testing gravity in the strong-field and highly relativistic regime~\cite{TheLIGOScientific:2016src,Abbott:2018lct,LIGOScientific:2019fpa,Abbott:2020jks,LIGOScientific:2021sio}, where new physics (e.g. deviations from GR) might emerge. Another compelling motivation for studying scalar-tensor theories is the inability of GR to explain the late-time acceleration of the Universe without introducing a cosmological constant or dark energy. Scalar-tensor theories can instead reproduce (at least potentially) cosmic acceleration with no additional dark energy or cosmological constant~\cite{Clifton:2011jh,Joyce:2014kja}, while still passing local tests of gravity in the solar system~\cite{Will:2014kxa}.

Among the various scalar-tensor theories, $K$-essence models have received significant attention. These theories, originally introduced in the context of inflation~\cite{ArmendarizPicon:1999rj,ArmendarizPicon:2000dh}, are characterized by non-canonical kinetic terms for the scalar field (with first derivative self-interactions), allowing 
for self-accelerated cosmic solutions at late times~\cite{Chiba:1999ka,ArmendarizPicon:2000dh}.
In more detail, the general K-essence action is given by:
\begin{align}\label{action}
S =& \int \mathrm{d}^{4}x \sqrt{-g}\left[\frac{M_{\mathrm{Pl}}^2}{2}R + K(\varphi, X)\right]\nonumber\\ &+ S_m[A(\varphi,X) g_{\mu\nu}, \Psi_m],
\end{align}
where $M_{\mathrm{Pl}}=(8\pi G)^{-1/2}$ is the Planck mass, $R$ is the Ricci scalar 
for the Einstein-frame metric $g_{\mu\nu}$,  $X = g^{\mu\nu}\partial_\mu \varphi \partial_\nu \varphi$ represents the kinetic term of the scalar field $\varphi$,  and $K(\varphi, X)$ is a general non-linear function that governs the dynamics. The matter fields $\Psi_m$
couple to the Jordan-frame metric $\tilde{g}_{\mu\nu}=A(\varphi,X) g_{\mu\nu}$,
where $A(\varphi,X)$ is the conformal factor. 

Another important feature of some $K$-essence theories is their ability to implement kinetic screening mechanisms, or $k$-mouflage~\cite{Babichev:2009ee}. The latter suppress deviations from GR on local scales, at least in quasi-static and quasi-spherical configurations~\cite{Babichev:2009ee,Kuntz:2019plo,terHaar:2020xxb,Bezares:2021yek,Shibata:2022gec,Lara:2022gof,Boskovic:2023dqk}. This  potentially allows for passing solar-system (and possibly also GW and binary-pulsar~\cite{Damour:1991rd,Kramer:2006nb,Freire:2012mg}) tests of gravity,  while
retaining significant modifications on cosmological scales,
where $K$-essence reduces to a Fierz-Jordan-Brans-Dicke (FJBD)~\cite{Fierz:1956zz,Jordan:1959eg,Brans:1961sx}
theory with a large conformal coupling.

K-essence models also remain viable in light of constraints from the propagation of GWs from GW170817~\cite{Monitor:2017mdv,TheLIGOScientific:2017qsa}, which 
requires
that the GW speed does not deviate significantly from the speed of light. Indeed, the latter bound, when combined with additional constraints on generic scalar tensor
theories coming from requiring the
absence of GW decay into dark energy  \cite{Creminelli:2018xsv,Creminelli:2019nok}   and the non-linear stability of the propagating scalar mode \cite{Creminelli:2019kjy},  reduces the viable class of scalar-tensor theories to the action \eqref{action} alone~\cite{Lara:2022gof}.

Despite their potential, $K$-essence theories face several challenges, particularly with regard to the well-posedness of the Cauchy problem in the non-linear regime~\cite{Babichev:2007dw,Bernard:2019fjb,Bezares:2020wkn,terHaar:2020xxb,Bezares:2021yek,Bezares:2021dma,Lara:2021piy,Shibata:2022gec}. Numerical simulations have shown that for certain initial data, the evolution equations can change character from hyperbolic to parabolic or even elliptic, leading to a ``Tricomi-type'' breakdown of the initial value problem. In other situations, although the equations remain hyperbolic at all times, the characteristic speeds can diverge, making numerical simulations practically unfeasible (``Keldysh-type'' breakdown). These issues complicate the study of K-essence in realistic systems, such as binaries of compact objects, and impede systematic investigations of the efficiency of kinetic screening in strong-field highly dynamical situations.

Several approaches have been tried, with partial success, to address Tricomi- and Keldysh-type breakdowns of the Cauchy problem in $K$-essence. Ref. \cite{Bezares:2020wkn}, 
showed that a subclass of $K$-essence theories can never experience Tricomi-breakdown because the evolution equations remain hyperbolic at all times (see also~\cite{Babichev:2007dw}). Even theories in this subclass, however, can still give rise to divergent characteristic speeds (Keldysh-type breakdown) in relevant physical systems such as gravitational collapse~\cite{Bezares:2020wkn,terHaar:2020xxb,Bezares:2021yek} (see also \cite{Akhoury:2011hr,Leonard:2011ce,Gannouji:2020kas} ). These divergences can be avoided by completing $K$-essence in the ultraviolet~\cite{Lara:2021piy} (which however may not be possible for theories that allow for screening~\cite{Adams:2006sv}); by modifying (``fixing'') the evolution equations by introducing additional auxiliary fields, 
which satisfy dissipative evolution equations and whose effect on the dynamics is negligible on secular scales~\cite{Cayuso:2017iqc,Allwright:2018rut,Bezares:2021yek}; or by choosing judiciously the gauge~\cite{Bezares:2020wkn}, introducing a non-trivial shift (even in spherical symmetry) to control the characteristic speeds and keep them finite at all times~\cite{Bezares:2021dma}.

Despite these successes, which have allowed Refs.~\cite{Bezares:2021yek,Bezares:2021dma} to simulate
stellar collapse to a black hole and the merger of neutron-star binaries, the physical implications of these results for the efficiency of kinetic screening and the viability of the theory are still unclear.  Refs.~\cite{Bezares:2021yek,Bezares:2021dma} found that gravitational collapse breaks kinetic screening, i.e. large amounts of scalar radiation are emitted in the process, and are redshifted to the frequencies observable by the Laser Interferometer Space Antenna (LISA), or even lower. Therefore, gravitational collapse can give rise to potentially observable effects on gravitational interferometers, 
provided that the scalar has a conformal coupling  $\sim {\cal O}(1)$ to matter on large scales.\footnote{Gravitational interferometers are generally outside the screening radius of the 
binary, i.e. they are in a region where the conformal coupling may, in principle, be $\sim {\cal O}(1)$. Still, one may argue that the detector is within the screening radius of the Sun or the Earth.
It should be noticed, however, that the geometry of the screening region in the presence of multiple bodies is complicated because the theory is not linear, e.g. Ref.~\cite{Boskovic:2023dqk} showed that the region near a static binary's saddle point is not screened, while being formally within the binary's screening radius. This may be  relevant for space-borne detectors such as LISA. }
Similarly, kinetic screening is efficient at suppressing scalar radiation from the late inspiral of neutron stars~\cite{Bezares:2021dma}, but Ref.~\cite{Bezares:2021dma}
finds no clear suppression in the quadrupole scalar radiation. This would have strong implications for the viability of $K$-essence theories with screening, especially if it could be extrapolated back to the early inspiral of binary neutron stars, where
a partially screened (or unscreened) scalar quadrupole emission would strongly impact the dynamics via radiation reaction.
Indeed, the latter regime is probed with high accuracy by binary pulsars, but cannot be treated with the standard perturbative post-Newtonian techniques in $K$-essence theories with screening, due to the importance of non-linearities in the local dynamics~\cite{Kuntz:2019plo,Boskovic:2023dqk}.

These results for quadrupole scalar emission are still uncertain, for several reasons. First, Ref.~\cite{Bezares:2021dma} could only perform a handful of relatively short simulations, which due to computational limitations could not investigate the range of physical scales of cosmologically motivated $K$-essence theories. In the latter, the region where screening is efficient at suppressing deviations from GR in spherical symmetry has a size of $\sim 10^{11}$ km for a star with mass $\sim M_\odot$. This screening radius is obviously much larger than the separation of binary pulsars and merging neutron-star binaries and than 
the wavelength of gravitational perturbations. Not only is it impossible to fully simulate such a system due to resolution requirements, but also the dynamics of $K$-essence depends on ratios of these very different scales, which poses problems of machine accuracy. For these reasons,  Ref.~\cite{Bezares:2021dma} simulated systems for which the screening radius is only parametrically larger than the binary's initial separation.

Second, the absence of screening for the scalar quadrupole  emission observed by Ref.~\cite{Bezares:2021dma}  in neutron star binaries is not observed in the scalar emission from deformed isolated stars~\cite{Shibata:2022gec}. The two results are not per se in disagreement: not only do the two investigations deal with different sources (binary vs isolated neutron stars), 
but Ref.~\cite{Bezares:2021dma} also finds suppression of the quadrupole in the post-merger signal (c.f. their Fig. 3, where smaller strong coupling scales $\Lambda$ correspond to larger screening radii). However, it is suggestive that Ref.~\cite{Shibata:2022gec} finds the suppression of the quadrupole scalar emission to take place only when the star's screening radius is much larger than the wavelength of the scalar waves. 
(This was also found in Galileon scalar-tensor theories~\cite{deRham:2012fg,deRham:2012fw,chu_retarded_2013,andrews_galileon_2013,Dar:2018dra,Brax:2020ujo,deRham:2024xxb}.) This is a condition which is not verified at the beginning of the simulations of Ref.~\cite{Bezares:2021dma}, where the two quantities are comparable.

Motivated by these facts, in this paper we further investigate the dynamics of binary compact-object systems in $K$-essence theories with screening using 3+1 dimensional numerical relativity simulations. To simulate more realistic separations of scales (between the screening radius and the wavelength of the scalar waves), we restrict to the decoupling limit, i.e. we only evolve the scalar sector of the theory for systems of a black hole and a neutron star. 
The decoupling limit was already used for similar investigations of the binary dynamics in Galileon scalar tensor theories~\cite{deRham:2012fg,deRham:2012fw,chu_retarded_2013,andrews_galileon_2013,Dar:2018dra,Brax:2020ujo,deRham:2024xxb}, where it was found to allow for a larger range of scale separations than possible in the full-fledged simulations of  Ref.~\cite{Bezares:2021dma}. Moreover, the choice of black hole-neutron star binaries
 allows one to maximize the expected scalar radiation (which, at least in simple scalar-tensor theories like FJBD, should be strongest for these systems~\cite{Damour:1992we,Will:1989sk}), while simplifying the equations and the initial data (because black holes have no hairs in shift-symmetry $K$-essence theories~\cite{Hui:2012qt,Sotiriou:2013qea,Capuano:2023yyh}). Moreover, for these mixed-type binaries, we find that the decoupling limit equations admit an analytic treatment for the scalar dipole emission, which we use to validate our numerical results. This analytic solution relies on a matched asymptotic expansion between the near zone and far zone of the binary's spacetime, which allows one to rigorously compute the dipole emission without restrictions on the masses and separations of the compact objects.
 
 We find that the scalar dipole emission is suppressed (i.e. screened), but kinetic screening is less efficient for the scalar quadrupole emission, even when a clear separation of scales between the scalar emission's wavelength and the screening radius is present. 
Despite the simplifications inherent to our decoupling-limit treatment, our results seem to suggest that GW generation by binary systems may be used to place constraints on $K$-essence theories, and therefore on effective field theories of dark energy.

This paper is organized as follows. In Section~\ref{sec:theoretical_setup}, we define the scalar-tensor theory that we study, give its equations of motion and comment on a standard simplification of the two-body problem which consists in replacing matter objects with point-particles. In Section~\ref{sec:matched_asymptotic_expansions} we present our method to obtain an analytic solution for the dipole, based on matched asymptotic expansions. We then show that the dipole is screened. In Section~\ref{sec:numerical_sol} we present our numerical results. We first use the analytic dipole solution to confirm the code, and afterwards we obtain the scaling of the quadrupole numerically. Our conclusions are given in Section~\ref{sec:conclusions}. The appendices contains further technical details: in Appendix~\ref{app:solve_mode_functions} we analytically solve for the radial function of the dipole field; in Appendix~\ref{app:matching} we explicitly perform the matching of the dipole in the near zone; in Appendix~\ref{app:different_regimes} we check the scaling of the dipole in different regimes of interest; in Appendix~\ref{app:mattdep} we check that our choice for the profile of the neutron star does not affect our numerical simulations; finally, in Appendix~\ref{app:convergence} we check the convergence of the numerical code.
Throughout this paper, we use units in which $\hbar=c=1$. 

\section{Theoretical setup} \label{sec:theoretical_setup}

Following Refs.~\cite{terHaar:2020xxb,Bezares:2021yek,Shibata:2022gec,Kuntz:2019plo,Boskovic:2023dqk}, we focus here on a class of $K$-essence models in which the non-linear kinetic term $K$ only depends on $X$ (and not on $\varphi$) and the conformal coupling only depends on $\varphi$ (and not on $X$). Action \eqref{action} thus becomes
\begin{equation}\label{action2}
    S = \int \mathrm{d} ^4 x \; \sqrt{-g} \bigg[ \frac{\mpl^2}{2} R + K(X) \bigg] + S_m \big[ A(\varphi) g_{\mu \nu}\,.\Psi_m \big]\,.
\end{equation}
For $K(X)$ and $A(\varphi)$, we consider the Effective Field-Theory (EFT) expansions
\begin{align} \label{eq:KX}
    &K(X) = - \frac{X}{2} + \frac{\beta}{4 \Lambda^4} X^2 - \frac{\gamma}{8 \Lambda^8} X^3 + {\cal O}(X^4)\\
    &A(\varphi)= \exp\left[2\alpha \frac{\varphi}{M_{\mathrm{Pl}}}+{\cal O}(\varphi^2)\right]\,,
\end{align}
which we truncate at the orders shown to perform the simulations described in the following. Here, the coefficients
$\alpha$, $\beta$ and $\gamma$ are dimensionless and $\sim {\cal O}(1)$, while  $\Lambda$ is the strong-coupling scale of the effective field theory.
For the scalar to give a  dark energy-like phenomenology, one needs $\Lambda \sim (H_0 \mpl)^{1/2} \sim 2 \times 10^{-3}$ eV, where $H_0$ is the present Hubble rate. 

It should also be noted that Tricomi-type breakdown of the Cauchy problem are avoided by construction if the function $K(X)$ satisfies the hyperbolicity condition~\cite{Babichev:2007dw,Bezares:2020wkn}
\begin{equation}\label{condition}
    1+2 \frac{ K''(X) X}{K'(X)}>0\,.
\end{equation}
This imposes constraints on the couplings $\beta$ and $\gamma$. For instance, if   we set $\beta=0$,
this condition is 
satisfied irrespective of $X$ if $\gamma>0$, although other choices of $\beta$ and $\gamma$ are possible. In the following we will set $\beta=0$ and $\gamma=1$, although our results do not sensitively depend on this choice.Moreover, note that a given form of the function $K(X)$, and in particular a choice of specific values of $\beta$ and $\gamma$, is stable against radiative corrections in the non-linear regime where screening is important~\cite{deRham:2014wfa,Brax:2016jjt}. %\sout{\AK{From an EFT point of view, setting $\beta=0$ seems to indicate a fine-tuning of the model we consider. While this condition is required to numerically solve for the system's dynamics, the analytical calculation of the dipole that we will present can also be done in more generic models where $\beta \neq 0$. Apart from different numerical factors, we find that the dipole field is similarly screened in these models.   }}

By varying the action \eqref{action2}, the equation of motion for the scalar field is
\begin{equation} \label{eq:scalarEOM}
    \nabla_\mu \big[ K'(X) \nabla^\mu \varphi \big] = \Sigma~,
\end{equation}
where the source is 
\begin{equation}
    \Sigma=\frac12 \frac{1}{\sqrt{-g}} \frac{\delta S_m}{\delta \varphi}=\frac{\alpha}{2 \mpl} T~,
\end{equation}
with
$T = g^{\mu \nu}T_{\mu \nu}$  the trace of the matter energy-momentum tensor
${T}^{\mu\nu}=(2/\sqrt{-{g}}) ({\delta S_m}/{\delta {g}_{\mu\nu}})$ in the Einstein frame. 
Refs.~\cite{terHaar:2020xxb,Bezares:2021yek,Shibata:2022gec,Boskovic:2023dqk} performed full-fledged numerical relativity simulations of binary systems, evolving  the metric, the scalar field and a perfect fluid, which in their approach models neutron stars.

In the following, we will take a different, simplified approach. Following the tried-and-tested approach used in FJBD-like scalar-tensor theories~\cite{Will:1989sk,Damour:1992we} (see also Refs.~\cite{deRham:2012fg,deRham:2012fw,Dar:2018dra,deRham:2024xxb,Brax:2020ujo} for Galileon theories, or Ref.~\cite{Yagi:2013ava,Will:2018ont} for Lorentz-violating gravity), we will model binary systems by using two point particles, i.e. we describe matter by the \textit{effective} action
\begin{equation}\label{ppaction}
    S_m=- \sum_i \int \tilde{m}_i(\varphi) d\tilde{\tau}_i\,, 
\end{equation}
where $i$ runs from 1 to 2,   the proper times $\tilde{\tau}_i$ are defined with respect to the Jordan frame metric $\tilde{g}_{\mu\nu}=A(\varphi)g_{\mu\nu}$, i.e.
\begin{equation}
   d\tilde{\tau}=e^{ \frac{\alpha\varphi}{M_{\mathrm{Pl}}}}\sqrt{-g_{\mu\nu}d x^\mu d x^\nu}\,,
\end{equation}
and the Jordan-frame masses as written as functions of $\varphi$ to account for possible violations of the strong equivalence principle. In the literature~\cite{1975ApJ...196L..59E,Will:1989sk,Damour:1992we,Kuntz:2024jxo}, the ``sensitivities''
are defined as 
\begin{equation}
    s_i\equiv -\frac{\partial \ln \tilde{m}_i}{\partial \ln A}\,.
\end{equation}

The  source $\Sigma$ defined above then becomes
\begin{align}\label{sigmadef}
    \Sigma&=\frac12 \frac{1}{\sqrt{-g}} \frac{\delta S_m}{\delta \varphi} \nonumber\\
    &=- \frac{\alpha}{2 M_{\mathrm{Pl}}}  \sum_i \frac{m_i (1-2s_i) }{\sqrt{-g}u^t}\delta^{(3)}(\boldsymbol{x}-\boldsymbol{x}_i(t))  \,,
\end{align}
where $m_i\equiv \tilde{m}_i \exp{\alpha \varphi/ M_{\mathrm{Pl}}}$ are the Einstein-frame masses,  $u^\mu$ is the Einstein-frame four velocity, and $\boldsymbol{x}_i(t)$ are the spatial
coordinates of the $i$-th body as function of coordinate time.

The combinations ${\alpha}_i=\alpha (1-2s_i)$
are often referred to as ``scalar charges''~\cite{Damour:1992we}. In $k$-essence, because of the shift symmetry of the theory in vacuum, black hole solutions match the FJBD ones~\cite{Hui:2012qt,Sotiriou:2013qea,Capuano:2023yyh}), which 
have ${\alpha}=0$ (or equivalently $s=1/2$)~\cite{Will:1989sk,Damour:1992we}. 
For weakly gravitating stars, $s\approx0 $ in both 
$k$-essence and FJBD. However, in FJBD the sensitivities can be $s=O(0.1)$ for neutron stars. In $k$-essence, because local kinetic screening is efficient at suppressing scalar effects near compact stars~\cite{terHaar:2020xxb,Bezares:2021yek}, it is 
 reasonable to assume $s_i\approx 0$ also for neutron stars.

However, in the following, we will keep the sensitivities of stars generic and
non-vanishing. We will therefore parametrize our results in terms of the neutron star scalar charge ${\alpha}_{\rm NS}=\alpha (1-2s_{\rm NS})$, i.e. we will solve the scalar-field equation \eqref{eq:scalarEOM}
with the source
\begin{align}\label{sigmaused}
    \Sigma=- \frac{\alpha_{\rm NS}}{2 M_{\mathrm{Pl}}}  \frac{m_{\rm NS} }{\sqrt{-g}u^t}\delta^{(3)}(\boldsymbol{x}-\boldsymbol{x}_{\rm NS}(t))  \,,
\end{align}
where $m_{\rm NS}$ is the mass of the neutron star. Note that the mass of the black hole, $m_\mathrm{BH}$, does not appear explicitly, as it is multiplied by the (vanishing) black hole scalar charge in  \eqref{eq:scalarEOM}. However, the black hole is, of course, crucial to keep the neutron star in a bound orbit. In more detail, we will write the source $\Sigma$ in the center-of-mass frame of a circular binary consisting of a neutron star and a black hole. The angular frequency of the binary, $\Omega$, is related to the separation $a$ by Kepler's law, i.e.
\begin{equation} \label{eq:defOmega}
    \Omega = \sqrt{\frac{G(m_\mathrm{BH}+m_\mathrm{NS})}{a^3}}\,.
\end{equation}
Because, at least at leading order, the effect of gravity is already accounted for via Kepler's law, we will also set the metric $g_{\mu\nu}$ appearing in Eqs.~\eqref{eq:scalarEOM} and \eqref{sigmaused} to flat space.

\section{Matched asymptotic expansions: an analytic solution to the one-body problem} \label{sec:matched_asymptotic_expansions}

From the analytic side, it is  difficult to solve the equation of motion~\eqref{eq:scalarEOM} for the scalar because of its non-linear nature, which is crucial for screening (i.e.~one cannot solve the dynamics perturbatively in the screening region). The state-of-the-art computations of the scalar field generated by isolated bodies or binaries in $K$-essence can be summarized as follows:
\begin{itemize}
    \item \textit{Isolated bodies}: The spherical and static case was considered in Refs.~\cite{Babichev:2009ee,terHaar:2020xxb,Bezares:2021yek,Lara:2022gof}. Refs.~\cite{Bezares:2021yek,Shibata:2022gec} also considered oscillated stars. These systems show evidence for kinetic screening.     
    \item \textit{Binaries}: Already for a two-body {\it static} system in the decoupling limit,  there is no analytic expression for the two-body potential in $K$-essence theories.  Refs.~\cite{Kuntz:2019plo,Boskovic:2023dqk} studied this static problem semi-analytically and numerically, and found the screening to be present, although its efficiency is slightly reduced compared to isolated bodies, especially in the region near the binary's saddle point~\cite{Boskovic:2023dqk}. As mentioned in the introduction, Ref.~\cite{Bezares:2021dma} performed 
     full-fledged numerical relativity simulations, whose results we are aiming to extend with this paper.
\end{itemize}

     Besides this work in $K$-essence, there exists a significant body of work on binaries in Galileon theories with Vainshtein screening~\cite{deRham:2012fg,deRham:2012fw,chu_retarded_2013,andrews_galileon_2013,Dar:2018dra,Brax:2020ujo,deRham:2024xxb}.     
     While the treatment of binaries is approximate -- assuming not only the decoupling limit, but also extreme mass ratios, making it similar in spirit to the oscillating stars of Refs.~\cite{Bezares:2021yek,Shibata:2022gec} -- these  results show that scalar waves should be screened, if the wavelength of the corresponding radiation is smaller than the screening radius.

For comparable-mass binaries, the problem of analytically solving for scalar emission is still open. Refs.~\cite{deRham:2012fw,deRham:2012fg} proposed to deal with these systems 
in Galileon theories by linearizing equations of motion over a static background field generated by a fictitious particle located at the center-of-mass of the binary. While the idea is interesting, this description cannot  account for
the details of the field's dynamics in the near zone
(i.e. the region, of size comparable to the radiation's wavelength, where retardation effects are negligible).
In fact, as shown explicitly in Ref.~\cite{Boskovic:2018lkj}, screening becomes ineffective at the saddle point of a binary system. For comparable-mass binaries, this corresponds to the
the center of mass. This effect is not accounted for in the framework of Refs.~\cite{deRham:2012fw,deRham:2012fg}.

In this work, we will adopt an analytic approach similar to Refs.~\cite{deRham:2012fw,deRham:2012fw}, while at the same time consistently solving for the field dynamics in the near zone of a black hole-neutron star system. We will use the formalism of matched asymptotic expansions, a tool   used with success in the traditional post-Newtonian approach to the two-body problem in GR~\cite{Blanchet:2013haa}. By ``stitching'' together two solutions in a buffer zone $a \ll r \ll 2 \pi/\Omega$ (where $a$ is the separation of the two objects and $\Omega$ is the Kepler frequency defined in Eq.~\eqref{eq:defOmega}), we will consistently obtain a solution for the dipole emission. The extension of this method to higher multipole orders and/or to neutron star binaries is tricky, and will be left to future work.

\subsection{Static one-body problem and near-zone solution} \label{sec:static1body}

\begin{figure}[h]  
    \centering  % Center the image
    \includegraphics[width=0.5\textwidth]{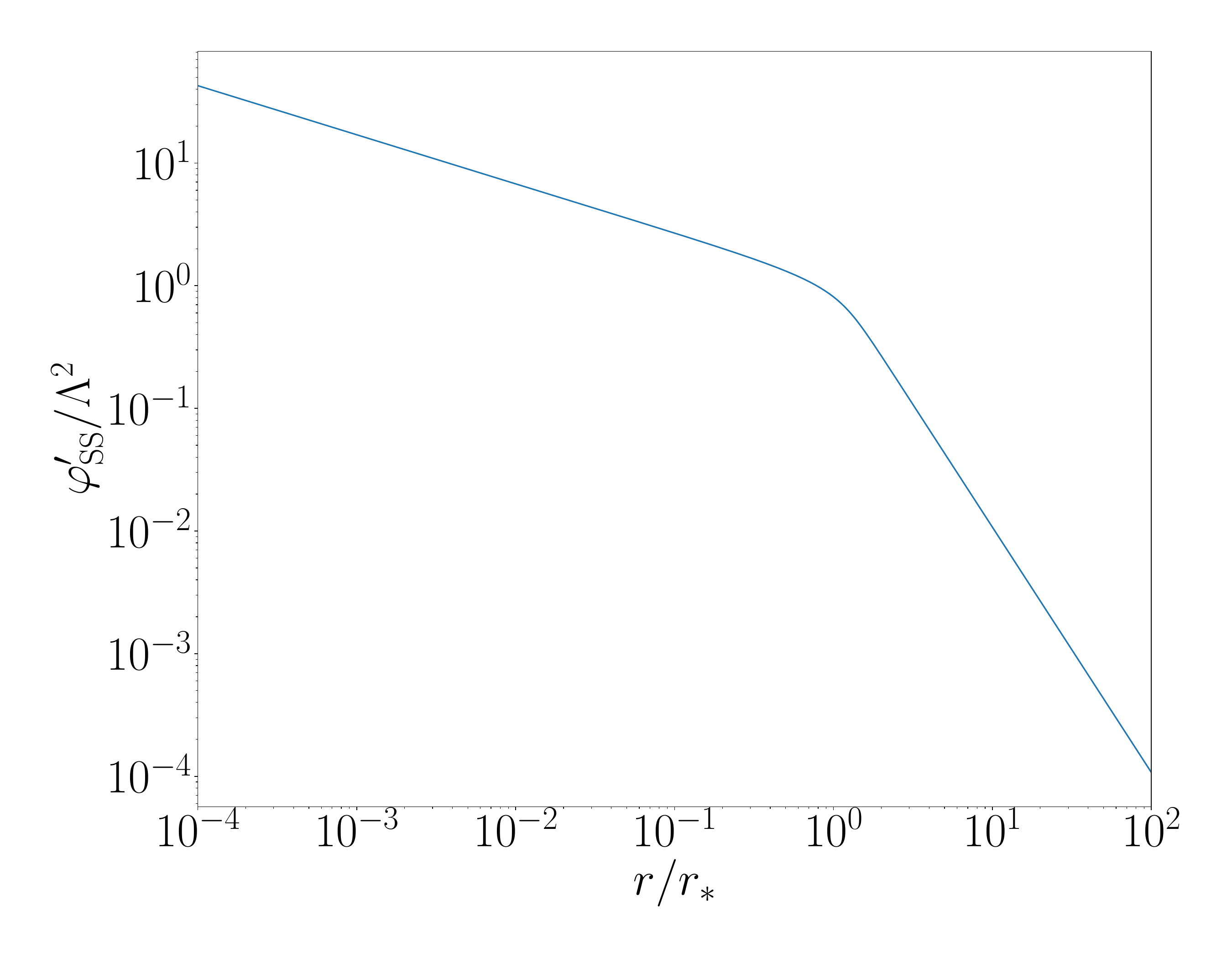}  % Adjust the width as needed
    \caption{Radial derivative of the spherically symmetric field for a single isolated neutron star of mass $m_\mathrm{NS}=1.4 \, M_\odot$ and with a scalar charge $\alpha_\mathrm{NS}=1$, obtained by solving Eq.~\eqref{eq:phiSS}. The strong-coupling scale was set to $\Lambda\approx 2\times 10^{-3}$ eV.  }  % Caption for the image 
    \label{fig:phiSS}  % Label for referencing the image
\end{figure}

Let us start by reviewing the solution to the field equations when the point-particle representing the neutron star is at rest~\cite{Babichev:2009ee,terHaar:2020xxb,Bezares:2021yek,Lara:2022gof}. By going to a spherical coordinate system centered on the neutron star, we can recast the scalar equation~\eqref{eq:scalarEOM} into the following ordinary differential equation
\begin{equation}
    \frac{1}{r^2} \partial_r \big( r^2 K'(\varphi_\mathrm{SS}'^2) \varphi_\mathrm{SS}' \big) = - \frac{\alpha_\mathrm{NS} m_\mathrm{NS}}{8 \pi r^2 \mpl} \delta \big( r \big) \; ,
\end{equation}
where $r = |\bm x - \bm x_\mathrm{NS} |$ is the distance to the neutron star, and $\varphi_\mathrm{SS}$ denotes the spherically symmetric field obtained in this static configuration. Integrating once, we obtain
\begin{equation}\label{eq:phiSS}
     K'(\varphi_\mathrm{SS}'^2) \varphi_\mathrm{SS}'  = - \frac{\alpha_\mathrm{NS} m_\mathrm{NS}}{8 \pi r^2 \mpl} \; .
\end{equation}
For our choice of $K(X)$, we cannot analytically integrate this equation a second time. However, we can get the asymptotic behavior of the field in two limits, corresponding to either the perturbative regime $K(X)\approx -X/2$ (where non-linear terms are negligible and screening is not active), or the non-linear regime in which $K(X)\approx -\gamma X^3/(8 \Lambda^8)$ (where non-linear terms dominate and give rise to screening). The two approximate solutions read
\begin{align}
\varphi_\mathrm{SS}(r) &= - \frac{\alpha_\mathrm{NS} m_\mathrm{NS}}{4 \pi \mpl r}  + \dots, \quad r \gg r_* \; , \label{eq:phiSSOutside} \\
&= C + \frac{5}{3} \left(  \frac{\alpha_\mathrm{NS} m_\mathrm{NS} \Lambda^8 r^3}{3 \pi \mpl } \right)^{1/5} + \dots, \quad r \ll r_* \; , \label{eq:phiSSInside}
\end{align}
where we have imposed that the scalar field vanishes at infinity, $C$ is a constant of integration, and $r_*$ is the nonlinear screening radius separating the two regimes. In order to find the precise value of $r_*$, we transform Eq.~\eqref{eq:phiSS} to canonical form, introducing the variable $y = (3/4 \Lambda^8)^{1/4} \varphi_{SS}'$:
\begin{equation} \label{eq:Y}
    (1+y^4)y = \bigg( \frac{r_*}{r} \bigg)^2 \; ,
\end{equation}
where
\begin{equation}\label{eq:defrs}
    r_* = \bigg( \frac{3}{4} \bigg)^{1/8} \frac{1}{\Lambda} \sqrt{\frac{\alpha_\mathrm{NS} m_\mathrm{NS}}{4 \pi \mpl}} \; .
\end{equation}
Using the cosmologically motivated value of the strong-coupling constant, $\Lambda \sim 2 \times 10^{-3}$ eV, one finds that for a typical neutron star $r_* \simeq 10^{11}$ km.
The full solution for $\varphi_\mathrm{SS}(r)$, obtained by numerically solving Eq.~\eqref{eq:Y}, is plotted in figure~\ref{fig:phiSS}. From this figure, it is obvious that screening heavily suppresses the value of the scalar on local scales~\cite{Babichev:2009ee,terHaar:2020xxb,Bezares:2021yek,Lara:2022gof}.

As pointed out in Ref.~\cite{Boskovic:2023dqk}, this isolated-star solution can be easily extended to represent a black hole-neutron star binary in the near zone, i.e. in the region (of size comparable to the wavelength of the emitted scalar waves, $|\bm x| \ll 2 \pi/ \Omega$) where retardation effects can be neglected. Indeed, in the near zone and at leading  order in $1/c$ (i.e. Newtonian order), one can neglect time derivatives $\partial_t/c$ in the scalar equation~\eqref{eq:scalarEOM}, which becomes 
simply
\begin{equation}\label{newt_lim}
    \partial_i (K'(X) \partial^i \varphi)=   - \frac{\alpha_{\rm NS} m_{\rm NS}}{2 M_{\mathrm{Pl}}} \delta^{(3)}(\boldsymbol{x}-\boldsymbol{x}_{\rm NS}(t)) \; ,
\end{equation}
where $X=\delta^{ij} \partial_i \varphi \partial_j \varphi$ and we have used the fact that $u^t=1+{\cal O}(1/c^2)$ and $g_{\mu\nu}=\eta_{\mu\nu}+{\cal O}(1/c^2)$. (Note that we use this flat-spacetime assumption exactly throughout this paper motivated by the decoupling limit, but even away from that limit, one can always set the metric to flat space in the near zone at leading Newtonian order~\cite{Will:1993ns,Boskovic:2023dqk}). 

It is then immediate to observe that a boost of the isolated star solution provides a solution to Eq.~\eqref{newt_lim}, i.e. in the near zone one has
\begin{equation} \label{eq:PNsol}
    \varphi(\bm x) = \varphi_\mathrm{SS}( | \bm x - \bm x_\mathrm{NS}(t) |)\,.
\end{equation}
Similarly, if one has a closed-form solution for a two-particle system, one could boost it to obtain a solution for a neutron star binary~\cite{Boskovic:2023dqk}. Unfortunately, static two particle solutions cannot be obtained analytically~\cite{Kuntz:2019plo,Boskovic:2023dqk}.

\subsection{Multipole expansion of the large-distance field} \label{sec:multipole}

Let us now consider the (vacuum) region at large distances from the source $| \bm x| \gg a$. We will attempt to solve the scalar equation \eqref{eq:scalarEOM} by modeling the scalar profile with a 
single-particle static solution located at the origin of the coordinates, plus a small time-dependent perturbation:
\begin{equation} \label{eq:multipolar_expansion}
    \varphi (\mathbf{x}) = \varphi_\mathrm{SS} (|\mathbf{x}|) +  \varphi_1(t,\mathbf{x})\,.
\end{equation}
We will solve for the field $\varphi_1$ by linearizing the equation, and we will check {\it a posteriori} the self-consistency of the method, i.e. we will determine the minimum distance at which $\varphi_1 \ll \varphi_\mathrm{SS}$. For more details, see Appendix~\ref{app:different_regimes}, where we will find this distance to be  the binary's separation $a$.

%imagine we look to the field at large distances from the source, so that we have EOM in vacuum. If we are sufficiently far away we expect on physical grounds to have a field very close to the spherically symmetric solution generated by a total scalar charge $\alpha_\mathrm{NS} m_\mathrm{NS}$ at the origin of the coordinates. We will make the meaning of ``far away'' more precise later on; for the moment let us just note that, for a Brans-Dicke field, ``far away'' means $| \bm x| \gg a$ (where $a$ is the radius of the orbit) so that we can perform the usual multipolar expansion of the field. Let us for the moment just assume that the field takes the form:
%$\begin{equation} \label{eq:multipolar_expansion}
%    \varphi (\mathbf{x}) = \varphi_\mathrm{SS} (|\mathbf{x}|) +  \varphi_1(\mathbf{x})
%\end{equation}
%where $\varphi_1 \ll \varphi_\mathrm{SS}$ at large distances. We will solve for the field $\varphi_1$ by linearizing the equations and we will check afterwards the self-consitency of the method, i.e. we will determine the limiting distance at which $\varphi_1 \ll \varphi_\mathrm{SS}$. We will find that this distance will be $a$, exactly like the Brans-Dicke case.

The linearized equation for  $\varphi_1$ reads
\begin{equation} %\label{eq:multipolar_expansion}
    \partial_\mu \bigg[ K'(X_\mathrm{SS}) \partial^\mu \varphi_1 +2 \partial^\mu \varphi_\mathrm{SS} K''(X_\mathrm{SS}) \partial_\nu \varphi_\mathrm{SS} \partial^\nu \varphi_1  \bigg] = 0\,,
\end{equation}
where $X_\mathrm{SS} = (\varphi_\mathrm{SS}')^2$. We recall that the right-hand side is zero because we are solving the scalar equation in vacuum far from the source. %We want to solve this equation with purely outgoing boundary condition at infinity. 
% In spherical coordinates we get
% \begin{align}
%     &\frac{1}{r^2} \partial_r \big[ r^2(K' + 2 X_\mathrm{SS}K'') \partial_r \varphi_1  \big] + K' \bigg[ - \partial_t^2 \varphi_1 \\
%     & + \frac{1}{r^2 \sin \theta} \partial_\theta \big( \sin \theta \partial_\theta \varphi_1 \big) + \frac{1}{r^2 \sin^2 \theta} \partial_\phi^2 \varphi_1 \bigg] = 0
% \end{align}
Because this equation is linear, we can expand $\varphi_1$ in  Fourier modes and  spherical harmonics:
\begin{equation}\label{eq:multipole_decomposition}
    \varphi_1 = \int \mathrm{d}\omega e^{-i \omega t} \sum_{\ell,m}  \mathcal{A}_{\ell m \omega} R_{\ell m \omega}(r) Y_{\ell m}(\theta, \phi) + \mathrm{c.c.}
\end{equation}
where ``c.c.'' denotes  complex conjugation, and $ \mathcal{A}_{\ell m \omega}$ is an amplitude which is arbitrary at the moment.
The differential equation for the radial function $R_{\ell m \omega}$ (which in the following we will refer to simply as $R$ to avoid clutter) is
\begin{equation} \label{eq:EOMR}
    \frac{1}{r^2} \partial_r \big[ r^2(K' + 2 X_\mathrm{SS}K'') R'  \big] + K' \bigg( \omega^2 - \frac{\ell(\ell+1)}{r^2}\bigg) R = 0\,.
\end{equation}
% This equation has two homogeneous solutions $R^{(1)}$ and $R^{(2)}$. 
% % that we normalize such that the Wronskian is
% % \begin{equation}
% %     R^{(1)} (R^{(2)})' - (R^{(1)})' R^{(2)} = \frac{1}{r^2(K' + 2 X_\mathrm{SS}K'')}
% % \end{equation}
% $R^{(1)}$ can be chosen to obey purely outgoing boundary conditions as large $r$, and $R^{(2)}$ purely ingoing conditions (i.e.  $R^{(1)} \sim e^{i \omega r}$ and $R^{(2)} \sim e^{-i \omega r}$).
% These boundary conditions 
% satisfy the scalar equation~\eqref{eq:EOMR} at large $r$ (outside  the screening radius).
%
We solve this equation with outgoing boundary conditions, i.e. $R \simeq e^{i \omega r}$ for $r \rightarrow \infty$, which of course satisfy the scalar equation~\eqref{eq:EOMR} at large $r$ (outside  the screening radius).
Given the complexity of the background scalar profile $\varphi_\mathrm{SS}$, we could not obtain  explicit analytic solutions to Eq.~\eqref{eq:EOMR} valid at all  $r$. However, it is possible to obtain an approximate solution by matching together two limiting regimes. We present the resulting solution for the radial function $R$  explicitly  in App.~\ref{app:solve_mode_functions}.

% The final solution for the field's perturbation $\varphi_1$ is therefore
% \begin{equation}\label{eq:phiLarger}
%     \varphi_1 = \int \mathrm{d}\omega e^{-i \omega t} \sum_{\ell,m} \mathcal{A}_{\ell m \omega} R^{(1)}_{\ell m \omega}(r) Y_{\ell m}(\theta, \phi) + \mathrm{c.c.}
% \end{equation}
% where $\mathcal{A}_{\ell m \omega}$ is an arbitrary amplitude. 

To determine the amplitude $\mathcal{A}_{\ell m \omega}$, we can match the solution for the field at large distances obtained in this Section to the near-zone solution of Sec.~\ref{sec:static1body}. The matching is performed 
explicitly in App.~\ref{app:matching} for the dipole mode $\ell=m=1$, and yields
$\mathcal{A}_{11\omega} =  \delta(\omega - \Omega) \tilde{\mathcal{A}}_{1 1 \Omega}$, with
\begin{align}
     \tilde{\mathcal{A}}_{11 \Omega} &= i  e^{ 2i \pi/5} \sqrt{\frac{10}{3}} \frac{\alpha_\mathrm{NS} m_\mathrm{NS}}{4 \Gamma(7/10) \mpl} \nonumber \\
     &\times r_*^{-4/5} \bigg( \frac{\Omega}{2 \sqrt{5}} \bigg)^{7/10}  \frac{m_\mathrm{BH}}{m_\mathrm{BH}+m_\mathrm{NS}}  a\,.
\end{align}
It should be stressed that this matching procedure assumes implicitly
that there is a \textit{buffer zone} where both solutions are valid. Indeed,  by self-consistently solving the equations, we prove in App.~\ref{app:different_regimes} that the buffer zone is finite and corresponds to the range $a \ll |\bm x | \ll 2 \pi/ \Omega$.

%How can we determine it? For this we need to know the behavior of the field close to the source, which we already have from the post-Newtonian solution in Section~\ref{sec:static1body}. We can thus obtain the constant $\mathcal{A}_{\ell m \omega}$ by \textit{matching} the solution for the field at large distances obtained in this Section to the post-Newtonian solution of Section~\ref{sec:static1body} which we explicitly do in appendix~\ref{app:matching} for the dipole field. This implicitly assumes that there is a \textit{matching zone} where both solutions are valid; by self-consistently solving the equations, we prove in appendix~\ref{app:different_regimes} that the matching zone is the range $a \ll |\bm x | \ll 2 \pi/ \Omega$. 
%We only report here our result for the amplitude of the dipole field, which reads $\mathcal{A}_{11\omega} =  \delta(\omega - \Omega) \tilde{\mathcal{A}}_{1 1 \Omega}$ with
%\begin{align}
 %    \tilde{\mathcal{A}}_{11 \Omega} &= i  e^{ 2i \pi/5} \sqrt{\frac{10}{3}} \frac{\alpha_\mathrm{NS} m_\mathrm{NS}}{4 \Gamma(7/10) \mpl} \nonumber \\
%     &\times r_*^{-4/5} \bigg( \frac{\Omega}{2 \sqrt{5}} \bigg)^{7/10}  \frac{m_\mathrm{BH}}{m_\mathrm{BH}+m_\mathrm{NS}}  a
%\end{align}
%where we recall that $r_*$ is the screening radius for an isolated neutron star in eq.~\eqref{eq:defrs}.
%The constant $ \mathcal{A}_{1-1 \omega}$ takes the same expression, with the replacement $\delta(\omega - \Omega) \rightarrow - \delta(\omega + \Omega)$. 

\subsection{Screening of  dipole emission}

Our final solution for the dipole scalar field  is
\begin{equation} \label{eq:phi1}
    \varphi_1 =  \tilde{\mathcal{A}}_{11 \Omega} R_{1 1 \Omega}(r)   Y_{11}(\theta, \phi) e^{-i \Omega t}  + \mathrm{c.c.}\,,
\end{equation}
with the radial function $ R_{1 1 \Omega}(r)$ presented in  App.~\ref{app:solve_mode_functions}, c.f. Eqs.~\eqref{eq:R1WKB}-\eqref{eq:R1smallr}. 

\begin{figure}[h]  
    \centering  % Center the image
    \includegraphics[width=0.5\textwidth]{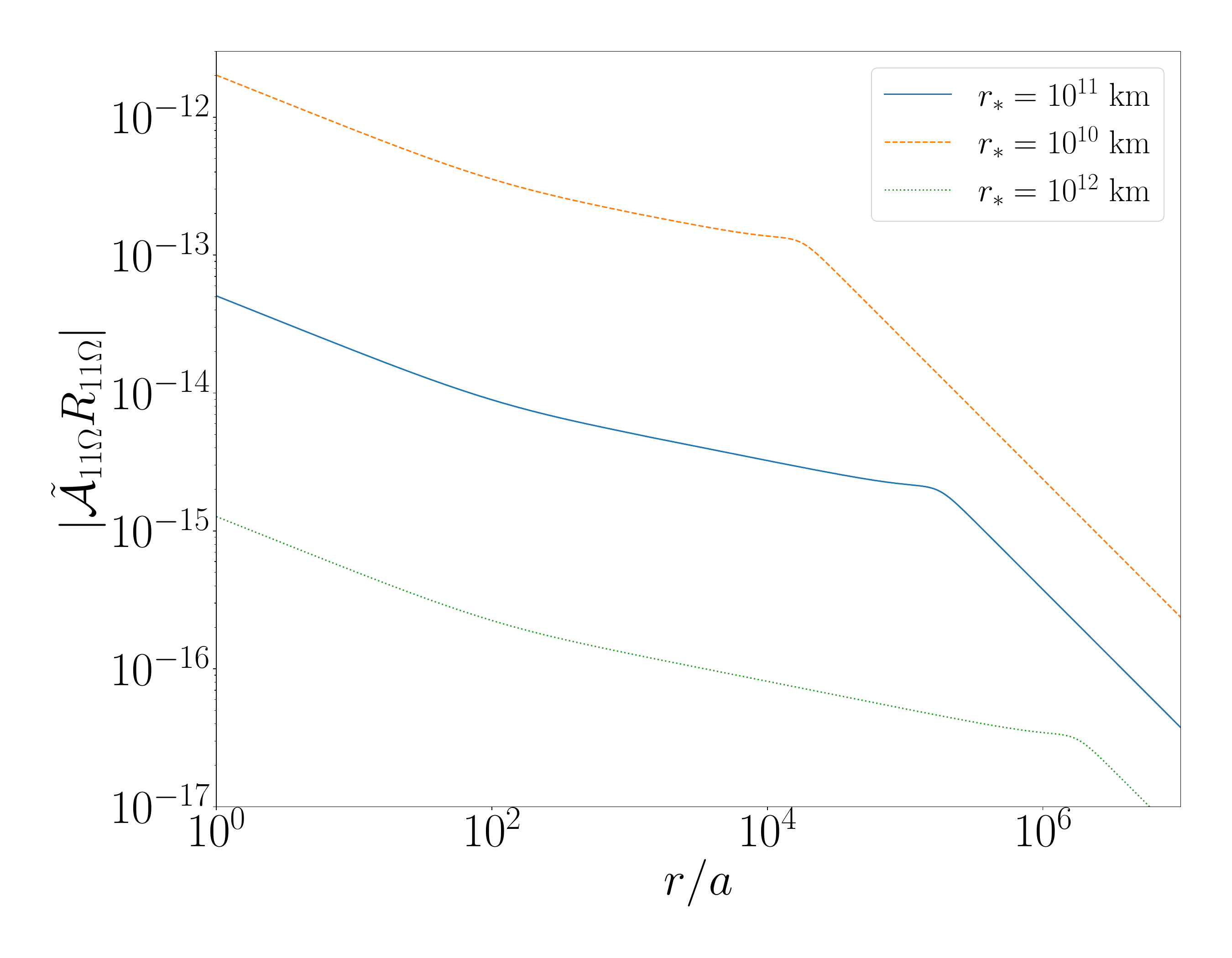}  % Adjust the width as needed
    \caption{Dipole scalar amplitude $\tilde{\mathcal{A}}_{11 \Omega}  R_{1 1 \Omega}$ as predicted by the analytic solution~\eqref{eq:phi1}, for different values of the screening radius. The parameters of the binary system are $m_\mathrm{NS}=1.4 \, M_\odot$, $m_\mathrm{BH}=50 \, M_\odot$, $a=10^4 (m_\mathrm{BH}+m_\mathrm{NS}) \simeq 8 \times 10^8$ m (so that $\Omega = (100 a)^{-1}$ from Eq.~\eqref{eq:defOmega}) and a scalar charge $\alpha_\mathrm{NS}=1$.    }  % Caption for the image
    \label{fig:dipole}  % Label for referencing the image
\end{figure}

In Fig.~\ref{fig:dipole}, we show the amplitude of the dipole predicted by Eq.~\eqref{eq:phi1}, for different values of the screening radius. 
The figure shows three different regimes for the scaling with $r$: $a \ll r \ll 2 \pi/\Omega$ ($\varphi_1 \propto r^{-2/5}$), $2 \pi/\Omega \ll r \ll r_*$ ($\varphi_1 \propto r^{-1/5}$) and $r \gg r_*$ ($\varphi_1 \propto r^{-1}$). These  approximate solutions are obtained in App.~\ref{app:different_regimes} by using explicit analytic expressions for the function $ R_{1 1 \Omega}(r)$ in these different regimes.  Figure~\ref{fig:dipole} also shows that the dipole emission is screened, as increasing $r_*$ suppresses the amplitude. 

More precisely, the scaling of the dipole field outside the screening radius relative to the single-particle solution $\varphi_\mathrm{SS}$ is
\begin{equation}
     \frac{\varphi_1}{\varphi_\mathrm{SS}} \sim v^{1/5} \bigg( \frac{a}{r_*} \bigg)^{4/5} \ll 1\,,
\end{equation}
c.f. App.~\ref{app:different_regimes}.
Besides confirming the validity of our perturbative expansion, this last estimate explicitly shows the screening of the dipole emission, since the ratio is proportional to the small factor $(a/r_*)^{4/5}$. 
This is in stark contrast with dipole emission in FJBD theory, where one finds simply $\varphi_1 / \varphi_\mathrm{SS} \sim v$. 
As a result, $K$-essence escapes the stringent bounds on scalar-tensor theories set by dipole scalar wave emission~\cite{Damour:1991rd,Kramer:2006nb,Freire:2012mg}. 

Before moving on, let us comment on higher-order terms in the multipole expansion. At quadrupole order, one needs to perturb the large-distance field to second order $\varphi = \varphi_\mathrm{SS} + \varphi_1 + \varphi_2$ with $\varphi_2 \ll \varphi_1$. Inserting this ansatz in Eq.~\eqref{eq:scalarEOM}, one obtains  a differential equation similar to Eq.~\eqref{eq:EOMR}, but with a source term (quadratic in the dipole perturbation $\varphi_1$) on the right-hand side. The solution to this  equation is more involved than in the dipole case, and will be left to future work.

\section{Numerical Simulations} \label{sec:numerical_sol}

In this section, we evolve neutron star-black hole binaries in the decoupling limit, by evolving Eq.~\eqref{eq:scalarEOM} with the source \eqref{sigmaused} in flat space. As we will see, this approach will allow us to simulate more realistic values of the strong coupling scale $\Lambda$ than possible with full-fledged numerical relativity simulations~\cite{Bezares:2021dma}.

%scenario by solving numerically the decoupling limit of the K-essence equation, \AK{which proves to be easier than solving the full equations}. That is, we will only numerically evolve the scalar field sector, while the matter describing the NS will be modeled with some smooth distribution prescribed to follow a Keplerian orbit in a flat spacetime background. The restriction to the decoupling limit is motivated by the pursuit of achieving realistic separations of scale where the wavelength of the scalar waves is smaller than the screening radius. 
We consider systems where the neutron star and the black hole have the same mass, $m_\mathrm{BH} = m_\mathrm{NS} = M$. While unrealistic from an astrophysical perspective, a light black hole mass yields a lower binary relative velocity for a fixed orbital frequency. This makes it easier to evolve numerically the equations for systems that satisfy the expected hierarchy of scales between the scalar perturbation's wavelength $\lambda$ and the screening radius $r_*$ ($\lambda\ll r_*$). Indeed, the mass of the black hole enters the simulations only through the Keplerian orbital frequency~\eqref{eq:defOmega}.

%The mass of the BH enters our equations only to relate orbital radius and frequency to the semimajor axis ($R_\mathrm{orbit} = a/2$ and $\Omega = \sqrt{G M/(4R_\mathrm{orbit}^3)}$ in the case we consider). 

 \subsection{Matter source}

When it comes to performing numerical simulations, the point particle approach is no longer viable. In order to model the matter source corresponding to the neutron star in the binary, we instead adopt a smooth  spherical profile to regularize the Dirac delta in Eq.~\eqref{sigmaused}, i.e. $\delta^{3}(\boldsymbol{x}) \to \rho(r)$ with

\begin{equation}\label{eq:matprofile} 
    \rho(r)=- \frac{A\,\tilde{r}^{2}}{(2\pi \sigma)^{3/2}}\exp(-\left(\frac{\tilde{r} - r_{p}}{\sigma}\right)^{2}),
\end{equation}
where $\tilde{r} =  \sqrt{x(t)^2 + y(t)^2 + z(t)^2}$,  being $x(t) = x + a/2 \cos{(\Omega t)}$, $y(t) = y + a/2 \sin{(\Omega t)}$ and $z(t) = z$, where we recall that $a$ is the radius entering Kepler's law in \eqref{eq:defOmega} and $A$ is a normalization constant ensuring that $\int 4\pi r^2 \rho \mathrm{d}r=1$. The value of $\sigma$ is fixed such that $99 \%$ of the mass is contained within a radius $r_s$, which, from now on, we will refer to as the radius of the star. 
This describes a smooth shell-like profile, and $r_{p}$ is a parameter that adjusts the shell's size. The factor of $\tilde{r}^2$ in front of the exponential is included to enforce $\rho$ to vanish at the center of the star. 
 
\begin{figure}[h]  
    \centering  % Center the image
    \includegraphics[width=0.475\textwidth]{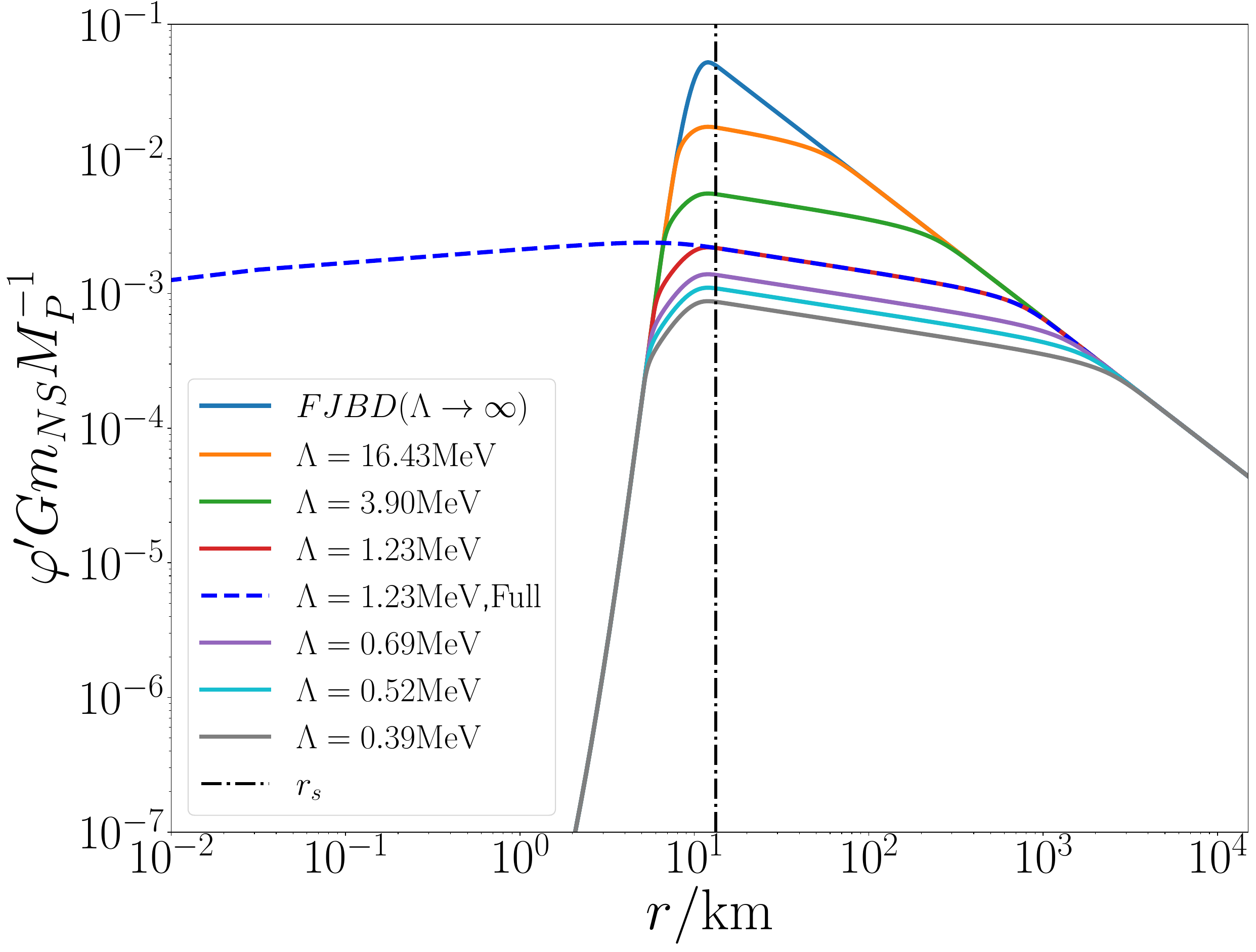}  % Adjust the width as needed
    \caption{The scalar field's radial derivative $\varphi'$  as a function of the distance $r$ from the center, for static spherical solutions with different values of $\Lambda$ and matter profiles. The dashed blue line corresponds to the solution for 
    a Gaussian profile for 
    $\Lambda=1.23 \text{ MeV}$, while the solid lines correspond to shell-like profiles. In all cases, the mass is set to $m_{\rm NS}=M_\odot$ and the radius to $r_s=13.3$ km, which is indicated with the vertical dot-dashed black line.}  % Caption for the image
    \label{fig:static_sols}  % Label for referencing the image
\end{figure}

This seemingly involved choice of the matter profile is useful to limit the resolution needed to model isolated-star screening solutions. To illustrate this point, Fig.~\ref{fig:static_sols} shows the static solutions for the shell-like profile for different values of $\Lambda$. As can be seen, when $\Lambda$ decreases, the scalar field gradient gets more and more suppressed as a result of screening. Also, the screening radius within which the suppression occurs increases as $\Lambda$ decreases, as predicted by Eq.~\eqref{eq:defrs}.
 
 Moreover, for fixed $\Lambda$, we  compare the solutions 
 obtained with the shell-like profile
of Eq.~\eqref{eq:matprofile} to those obtained 
with a Gaussian profile, i.e. by 
removing the $\tilde{r}^{2}$ term in  Eq.~\eqref{eq:matprofile} and setting $r_{p}=0$. In the latter case,   the derivative of the scalar field remains non-zero well inside the star, in contrast to the shell-like solutions. Note that regularity demands the gradient of the scalar field to vanish at the origin. 
Indeed, as shown in Ref.~\cite{terHaar:2020xxb}, the scalar field gradient does indeed go to zero linearly in $r$ near the origin, but only for very small $r$, c.f. their Fig.~1. 
In order to numerically resolve the scales at which the scalar field gradient becomes linear in $r$, a very fine spatial resolution is therefore required, which is restrictive in $3+1$ simulations. We thus avoid this problem altogether by using the shell-like profile in our simulations.

We stress that this choice does not affect our results and conclusions. Indeed, Fig.~\ref{fig:static_sols} shows that the exterior scalar profile is the same for the shell-like and Gaussian profiles. We show this in more detail in App.~\ref{app:mattdep}, where we look at the impact of the profile choice on dynamical quantities (such as outgoing radiation).

 \subsection{Initial conditions}

To maintain numerical stability during the evolution, especially as $\Lambda$ decreases, it is crucial to prescribe appropriate initial conditions for the scalar field $\varphi$. These can, in principle, be produced by boosting the shell-like solutions for $\varphi$ described in the previous section with the orbital velocity given 
by Eq.~\eqref{eq:defOmega}. As for
the scalar field derivative, one may try to set $\partial_t\varphi=0$ initially, letting the system relax and reach the physically realistic configuration
after an initial transient.

While this transient phase is generally not problematic, for high orbital frequencies $\Omega$ and small values of $\Lambda$, the evolution equations' characteristic structure can lead to catastrophic outcomes. Indeed, during the transient phase, the field can develop large characteristic speeds, causing a Keldysh-type breakdown of the Cauchy problem and requiring extremely restrictive Courant-Friedrichs-Lewy (CFL) conditions to keep the simulation feasible. These CFL conditions seem necessary throughout the simulation due to  ``shock-like'' features originating in the transient phase and persisting afterwards. To avoid these issues, we use the static solutions from the previous section as initial conditions for $\varphi$, but we boost them by prescribing a time-dependent orbital velocity  $\Omega(t)$ for the source, which slowly\footnote{ A similar approach was taken on \cite{Dar:2018dra,deRham:2024xxb}, where starting from a vacuum solution the non-linearities and the matter source are turned on slowly. However, we've found that such approach, although useful, tends to then demand restrictive CFL conditions through the evolution. The mentioned studies also find this and need  $\Delta_{t}/\Delta_{x} \sim 0.05$ and   $\Delta_{t}/\Delta_{x} \sim 0.005$ respectively. In contrast, we can carry out our simulations with one to two orders of magnitude less restrictive CFL factors.}  evolves from zero to its Keplerian value of Eq.~\eqref{eq:defOmega} on a timescale $t_{\rm{ramp}}=3.9\text{ ms}$:
 \[
\Omega(t) =
\begin{cases} 
\Omega\displaystyle{\frac{\left(2t_{\rm{ramp}} - t\right)\,t}{t_{\rm{ramp}}^{2}}} & \text{if } 0<t < t_{\rm{ramp}}, \\
\Omega & \text{if }  t \geq t_{\rm{ramp}}.
\end{cases}
\]
 \subsection{ Set-up and Numerical scheme}

For the simulations presented in this work, we
assume that neutron star sensitivities vanish, so that $\alpha_{\rm NS}=\alpha$, and we 
set the conformal coupling to  $\alpha =1/2$. We fix the mass of the star to one solar mass $m_{NS}=M_{\odot}$, its radius to $r_{s} = 13.3 \text{km}$, and the orbital separation to $a =52.8 \text{km}$. The values of $\Lambda$ are as high as $16.4 \text{MeV}$ and as low as $0.39\text{MeV}$, with $r_{*} \sim 2100 \text{km}$. We also simulate the FJBD case, corresponding to the limit $\Lambda \rightarrow \infty$.
 
To evolve Eq.~\eqref{eq:scalarEOM}, we reduce it to a first-order system by introducing the first-order variables $\Psi \equiv \partial_{t}\varphi$, $\psi_{x} \equiv \partial_{x}\varphi$, $\psi_{y} \equiv \partial_{y}\varphi$ and $\psi_{z} \equiv \partial_{z}\varphi$,
\begin{align}
    &\partial_{t}\left( K^{\prime}(X) \Psi\right) =  \partial_{i}\left(K^{\prime}(X) \psi_{i} \right) -\Sigma~,\\
    &\partial_{t}\psi_{i} = \partial_{i}\Psi~,\\
    &\partial_{t} \varphi = \Psi~.
\end{align} 
Our numerical code is a Simflowny-based high-performance 3D code  \cite{ARBONA20132321,ARBONA2018170,PALENZUELA2021107675,simflowny2021} and runs under the SAMRAI infrastructure \cite{https://doi.org/10.1002/cpe.652,GUNNEY201665,SAMRAIurl}, which provides parallelization and the adaptive mesh refinement (AMR) required to solve the different scales in the problem. We use fourth-order finite difference operators to discretize our equations and a fourth-order Runge-Kutta time integrator. Our computational domain corresponds to the range $[-12500,12500]^{3} \text{km}$ and contains 10 refinement levels. Each level has twice the resolution of the previous one, achieving a resolution of $\Delta x_{10}= 0.123 \text{km}$ on the finest grid. We use a Courant factor $\lambda_{c} \equiv \Delta t_{l}/\Delta x_{l} = 0.3$ on each refinement level $l$.

The large spatial domain is necessary to adequately extract the outgoing radiation at radii larger than the screening radius $r_{*}$. The response of a detector to scalar radiation is encoded in the Jordan-frame Newman-Penrose invariant $\phi_{22} $ \cite{Eardley:1973br}, which far from the source can be approximated \cite{Barausse:2012da,Bezares:2021yek} by
\begin{equation}
    \phi_{22} = -\alpha \sqrt{16\pi G}\partial_{t}^{2}\varphi + \mathcal{O}\left(1/r^{2}\right)\,.
\end{equation} 
This expression assumes that the scalar field decays as $\propto 1/r$, which is a good approximation at distances larger than the screening radius $r_{*}$. The strain $h_s$ produced by the scalar radiation is computed by integrating $\phi_{22}$ twice in time, i.e.  $\phi_{22} = \partial_{t}^{2}h_{s}$, which implies $h_s\propto\alpha \varphi$. Therefore, to study scalar emission, we extract $\varphi$ at extraction radii $\mathcal{R}_{E}=\{ 3692, 4429, 5168, 5900, 7383, 8860,10336,  11813\} \text{km}$, decompose it in spherical harmonics, and extrapolate its value to infinity. For this work, we will focus on the dominant contribution from the dipole ($l=m=1$), and the quadrupole $l=m=2$ component.

 \subsection{Results}

\begin{figure}[h]  
    \centering  % Center the image
    \includegraphics[width=0.475\textwidth]{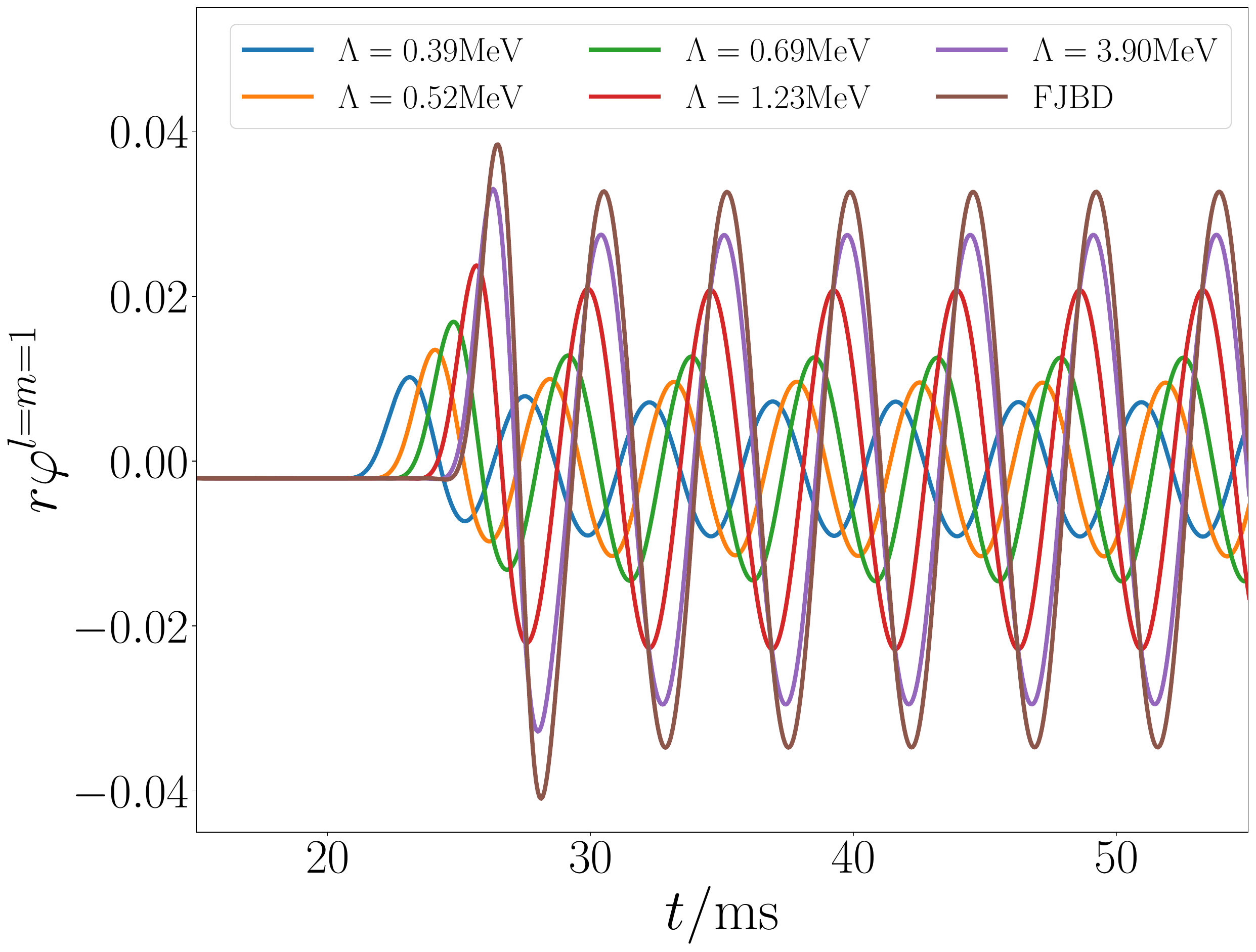}  % Adjust the width as needed
    \caption{Waveforms for the dipole $l=m=1$ mode of the outgoing scalar radiation, for different values of the strong coupling scale $\Lambda$, extracted at a radius of $r=7383 \mathrm{km}$.}  % Caption for the image
    \label{fig:waveform_dipole}  % Label for referencing the image
\end{figure}

\begin{figure}[h]  
    \centering  % Center the image
    \includegraphics[width=0.475\textwidth]{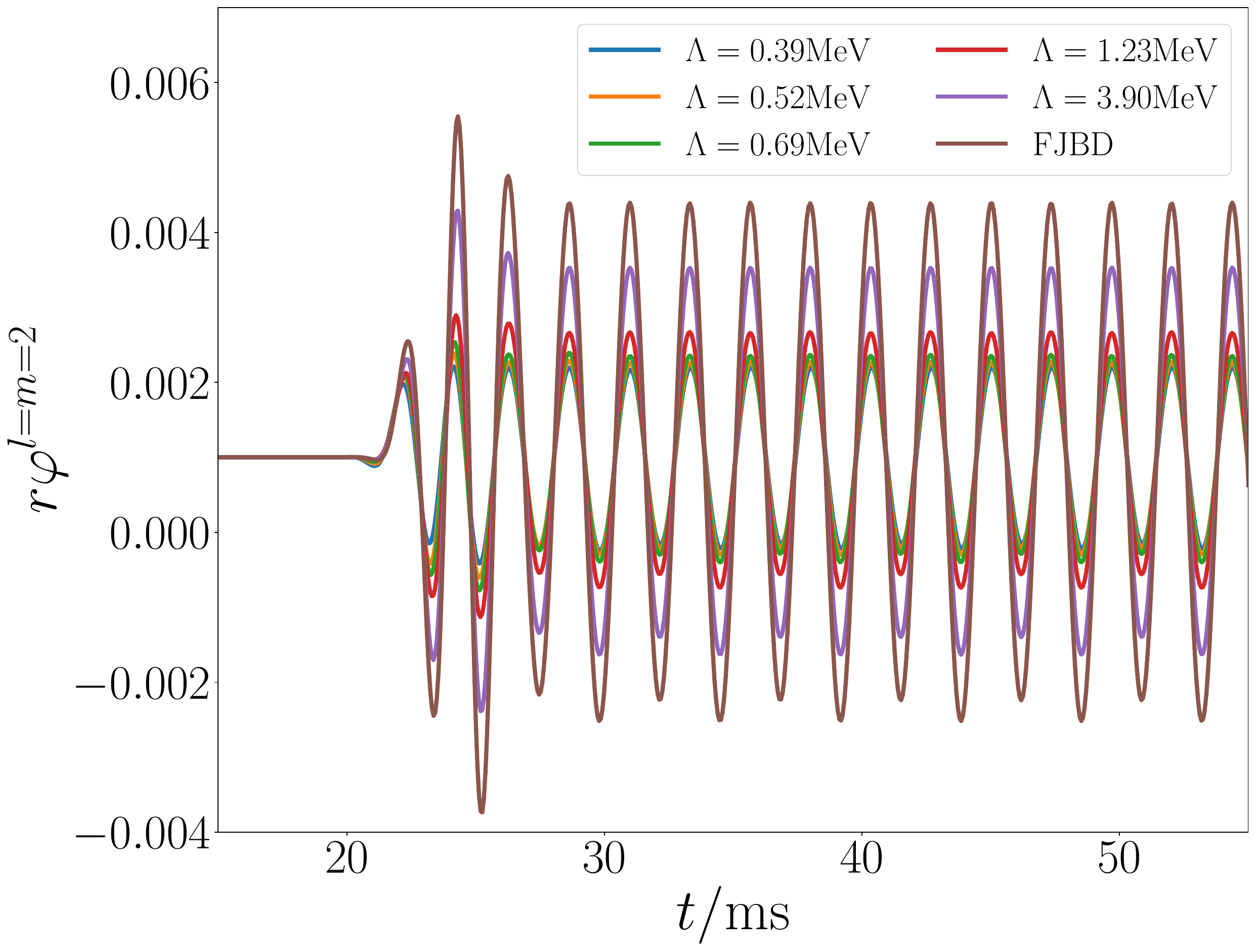}  % Adjust the width as needed
    \caption{Waveforms for the quadrupole $l=m=2$ mode of the outgoing scalar radiation, for different values of the strong coupling scale $\Lambda$ extracted at a radius of $r=7383 \mathrm{km}$. The first peak (after the transient phase) of the quadrupole waveforms have been aligned for ease of comparison.}  % Caption for the image
    \label{fig:waveform_quadrupole}  % Label for referencing the image
\end{figure}

With the setup discussed in the previous sections,  we performed simulations in a wide range of values for the strong coupling constant $\Lambda$. The purpose of these simulations is also to cover the regime where the wavelength of the outgoing radiation is smaller than the screening radius, i.e. $\lambda < r_{*}$. In Figs.~\ref{fig:waveform_dipole} and~\ref{fig:waveform_quadrupole}, we show the waveforms for the scalar  dipole $l=m=1$ and quadrupole $l=m=2$ modes, respectively. From these plots, it is clear that as the value of $\Lambda$ decreases, the screening mechanism suppresses the amplitude of the radiation. 

\begin{figure}[h]  
    \centering  % Center the image
    \includegraphics[width=0.475\textwidth]{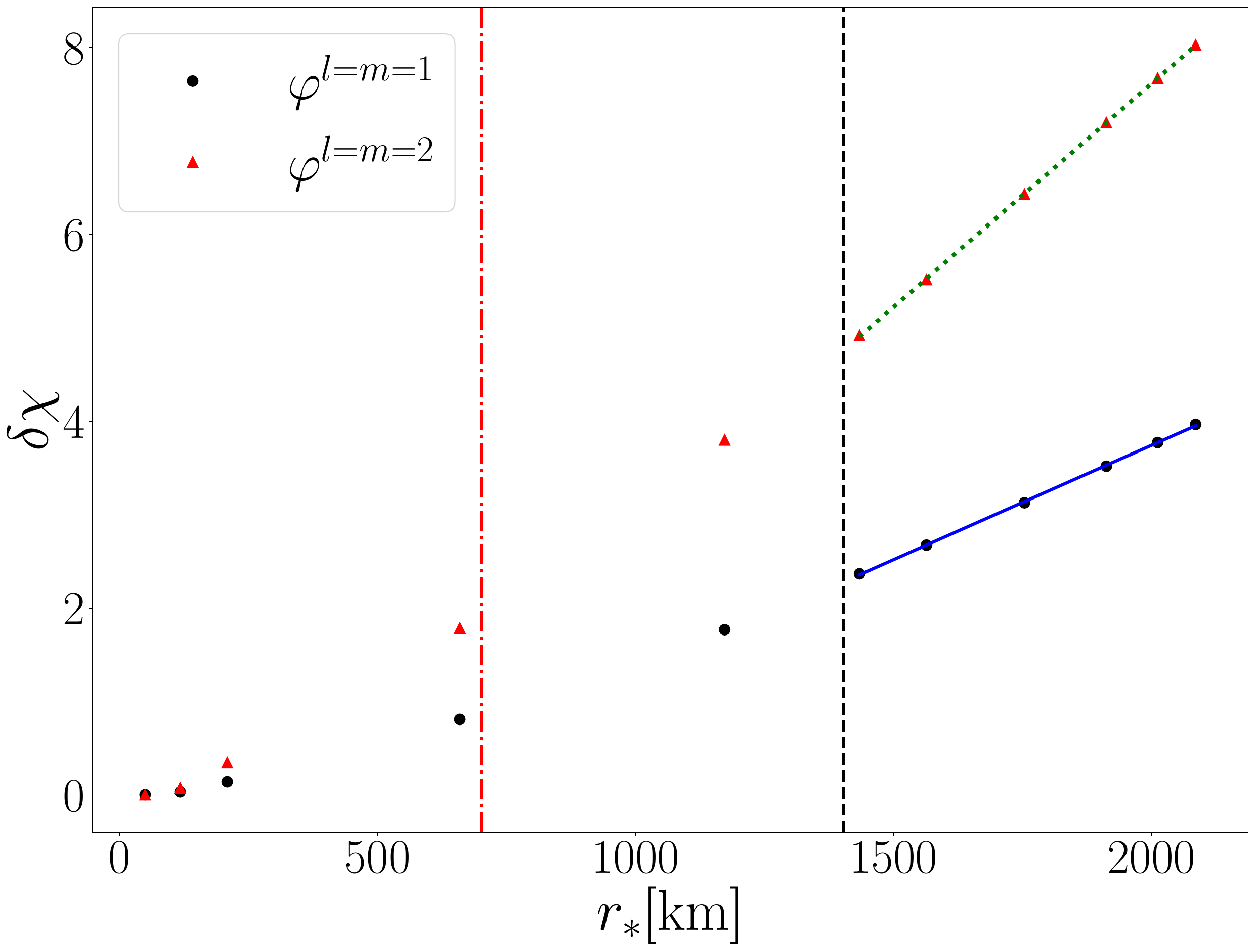}  % Adjust the width as needed
    \caption{Phase shift for the dipole $l=m=1$ and quadrupole $l=m=2$ waveforms as a function of the screening radius $r_*$. The blue solid and the green dotted lines correspond to linear fits to the last six points in the plots for the dipole and quadrupole, respectively. The values of $r_{*}$ on the right of the vertical black dashed and red dotted-dashed lines are those that are larger than the wavelength of the dipole and quadrupole radiation, respectively, i.e. is $\lambda_{l=m=1}<r_{*}$ and $\lambda_{l=m=2}<r_{*}$.   }  % Caption for the image
    \label{fig:phase}  % Label for referencing the image
\end{figure}

In the dipole radiation plot, we observe a clear phase shift for different values of $\Lambda$. This phase shift is also present in the quadrupole radiation, though larger, so we aligned the waveforms to provide a clearer view. The exact phase shift is found by fitting sinusoids to the waveforms, extracting the phase $\chi$, and calculating the difference relative to the FJBD case, $\delta\chi \equiv \chi_{\Lambda} - \chi_{FJBD}$. In Fig.~\ref{fig:phase}, we plot $\delta \chi$ for dipole and quadrupole radiation as a function of  $r_{*}$. According to Eq.~\eqref{eq:dipole_pred}, the phase shift should be  $\delta \chi = 0.56 \Omega r_{*}$ for the dipole\footnote{In more detail, for a fixed radius and using Eq.~\eqref{eq:dipole_pred} into 
 Eq.~\eqref{eq:phi1}, we obtain  $\varphi_1\propto \cos \Omega(t+0.56r_*)$ and thus a positive phase shift if $r_*$ increases. }.

To verify this in our simulations, we fit a linear function to  $\delta \chi$ for the values of $r_{*}$ satisfying $\lambda < r_{*}$. The fits yield  $
\delta \chi (\varphi^{l=m=1})= 0.55 \Omega r_{*} $ and $\delta \chi (\varphi^{l=m=2})= 1.07 \Omega r_{*} $, showing excellent agreement with the analytic prediction for the dipole phase shift. This also indicates that the quadrupole phase shift  
is roughly double that of the dipole, as expected given that the quadrupole radiation's frequency is $2 \Omega$.

\begin{figure}[h]  
    \centering  % Center the image
    \includegraphics[width=0.475\textwidth]{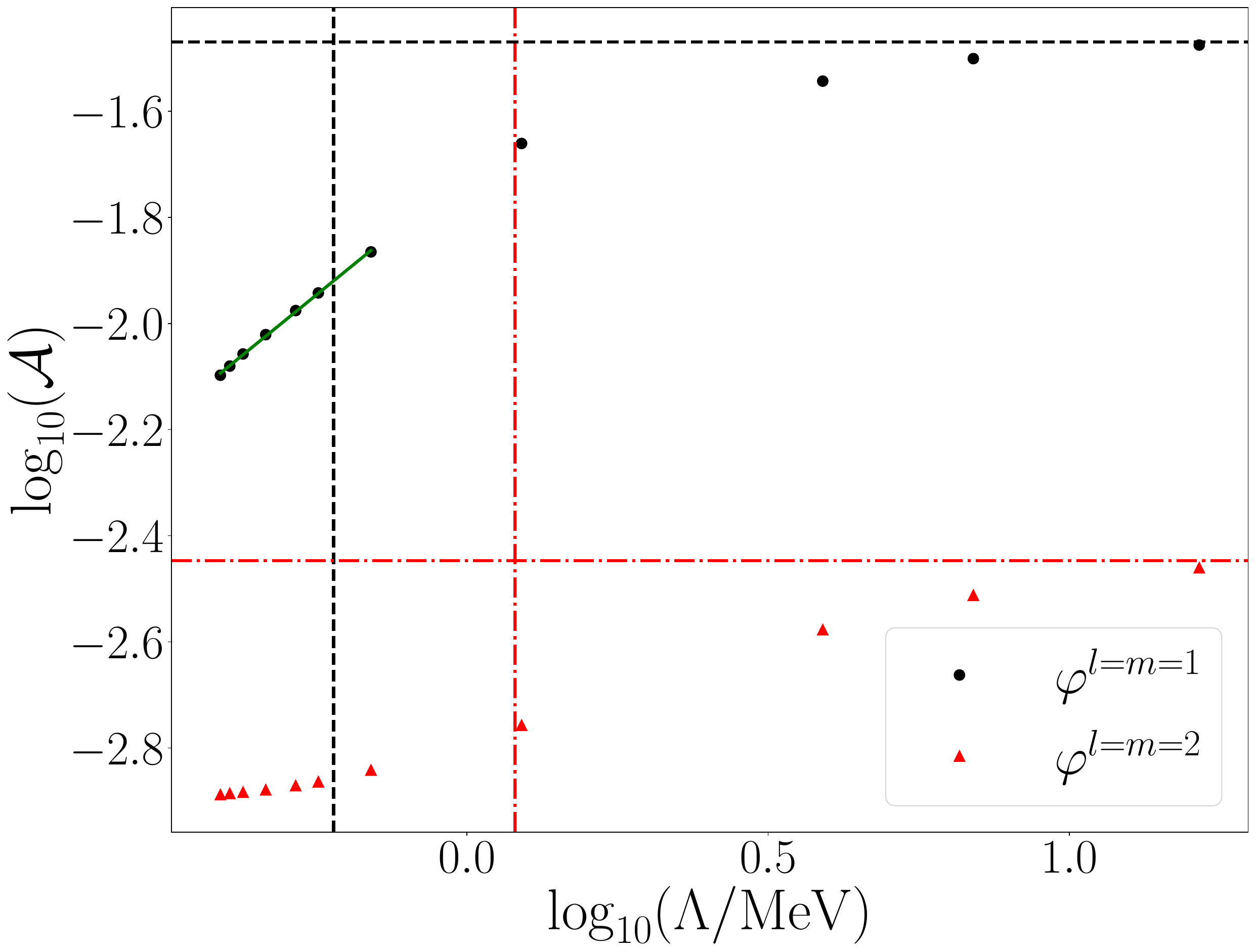}  % Adjust the width as needed
    \caption{Value of the amplitude $\mathcal{A}$ as a function of $\Lambda$ for the dipole and the quadrupole waveforms. The black horizontal dashed line and the red horizontal dotted-dashed line represent the value of $\mathcal{A}$ for the FJBD simulation for the dipole and quadrupole, respectively. Values of $\Lambda$ to the left of the black vertical dashed line and the red vertical dotted-dashed line represent the values for which $\lambda_{l=m=1} < r_{*}$ and $\lambda_{l=m=2} < r_{*}$ respectively. The solid green line corresponds to a linear fit to the dipole radiation curve, the slope of this fit corresponds to the exponent at which the amplitude is screened/damped as $\Lambda$ decreases.    }  % Caption for the image
    \label{fig:Amplitude}  % Label for referencing the image
\end{figure}

To study how the waveform amplitude depends on $\Lambda$, we fit sinusoids to several waveforms at different extraction radii and extract their amplitudes. The amplitude behaves as $A(r) = \mathcal{A}/r + \mathcal{O}(r^{-2})$ outside the screening radius. By fitting this form, we extract $\mathcal{A}$ for various values of $\Lambda$.
 Fig.~\ref{fig:Amplitude} shows the dependence of $\mathcal{A}$ on  $\Lambda$ for both dipole and quadrupole waveforms. The plot shows that $\mathcal{A}$ is suppressed compared to the FJBD case and decreases as $\Lambda$ decreases. This indicates that the screening mechanism effectively suppresses the outgoing dipole and quadrupole radiation. Focusing on the dipole in the region of the parameter space where $\lambda_{l=m=1} < r_{*}$, we can compare our numerical results to the analytic prediction of Eq.~\eqref{eq:dipole_pred}, i.e. $\mathcal{A} \propto r_{*}^{-4/5} \propto \Lambda^{4/5} $. Performing a linear fit to the dipole amplitudes in Fig.\ref{fig:Amplitude}, we find our numerical results to scale as $\mathcal{A} \propto \Lambda^{0.93}$, i.e. a power law with exponent differing from the analytic prediction by about 16\%. Indeed, screening of dipole radiation in the numerical solutions is stronger than analytically predicted.
 This difference could arise because the analytic calculation is limited to linear order in small fluctuations around a static background (cf. \eqref{eq:multipolar_expansion}), while higher-order perturbations might also affect the dipole and alter the scaling with $\Lambda$.

\begin{figure}[h]  
    \centering  % Center the image
    \includegraphics[width=0.475\textwidth]{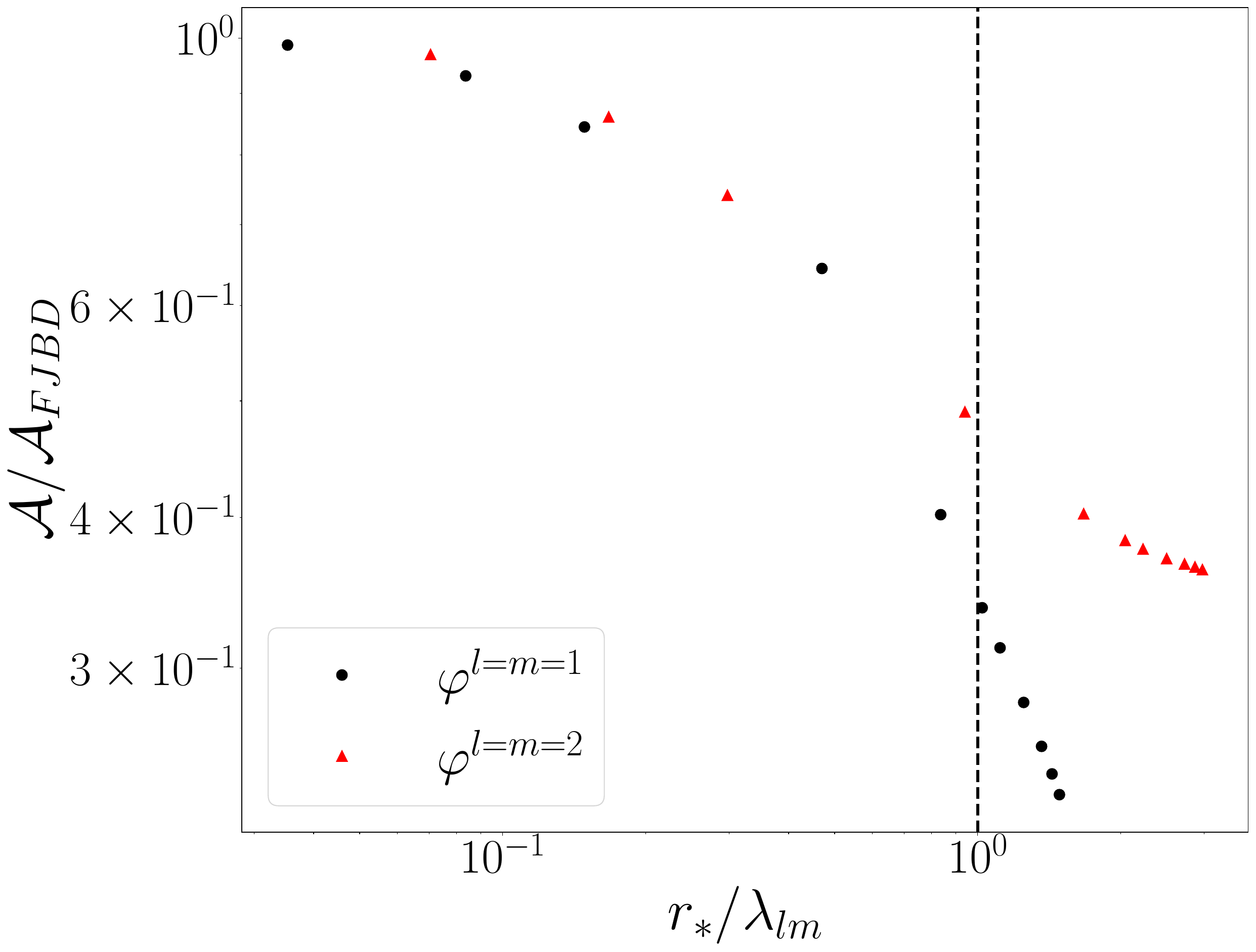}  % Adjust the width as needed
    \caption{Value of the amplitude $\mathcal{A}$ of the dipole and quadrupole relative to their value in FJBD $\mathcal{A}_{FJBD}$ as a function of $r_{*}/\lambda_{lm}$, where $\lambda_{lm}$ is the wavelength corresponding to $\varphi^{lm}$. The vertical dashed line indicates where $r_{*}/\lambda_{lm} = 1$.  }  % Caption for the image
    \label{fig:Amplitude_FJBD}  % Label for referencing the image
\end{figure}

We also plot, in Fig.~\ref{fig:Amplitude_FJBD}, the value of $\mathcal{A}$ relative to its corresponding value in FJBD $\mathcal{A}_{FJBD}$ as a function of $r_{*}/\lambda_{lm}$, where $\lambda_{lm}$ is the wavelength corresponding to $\varphi^{lm}$. Here, it is clear to see the scaling of the dipole and quadrupole resemble each other in the region where $r_{*}/\lambda_{lm}<1$. However, for $r_{*}/\lambda_{lm} \geq 1$ this pattern changes. For $r_*/\lambda_{lm} \sim 1$, an inflection point appears in the behavior of $\mathcal{A}$ for the quadrupole, i.e. the suppression of the quadrupole amplitude slows as $\Lambda$ decreases, almost flattening out at the lowest $\Lambda$ we can simulate. Comparing to the FJBD quadrupole amplitude, we see that the quadrupole is only suppressed by a factor $\lesssim 3$. 
This suggests that quadrupole radiation from a neutron star-black hole  binary may play a significant role in constraining $K$-essence theories with screening, although we are still unable to determine if the flattening of the quadrupole can be extrapolated to realistic values of the strong coupling scale ($\Lambda\sim10^{-3}\,$eV).

\section{Conclusions} \label{sec:conclusions}
In this work, we investigated the scalar emission from neutron star-black hole binaries in scalar-tensor theories with kinetic screening ($K$-essence theories). Our approach includes both 3+1 dimensional numerical relativity simulations in the decoupling limit and analytical predictions derived from matched asymptotic expansions.

Our simulations show that the scalar dipole emission is strongly suppressed by kinetic screening, especially as the strong coupling constant $\Lambda$ decreases and the screening radius grows larger than the wavelength of the emitted radiation. This suppression of dipole radiation agrees with analytical predictions, with phase shifts and amplitude reductions that match well within the expected range. 

The quadrupole scalar emission, however, shows a different behavior. While screening does initially suppress the quadrupole amplitude as $\Lambda$ decreases, like for the scalar dipole, the suppression eventually slows down when $\Lambda\to0$, even when the screening radius exceeds the radiation's wavelength. Indeed, the scalar quadrupole amplitude appears to asymptote to a constant as  $\Lambda\to0$, with an overall suppression at most by a factor $\sim3$ relative to quadrupole emission in FJBD theories at the lowest $\Lambda$ we can simulate.  This suggests that scalar quadrupole radiation could play a crucial role in constraining $K$-essence theories, particularly in the context of next-generation gravitational-wave detectors~\cite{Maggiore:2019uih,Evans:2021gyd}, which will observe mixed neutron star-black hole binaries in the early inspiral phase. Indeed, a quadrupole scalar emission that is only partially screened 
could speed up the phase evolution of mixed binaries due to the enhanced radiation reaction.

\acknowledgements

We thank Carlos Palenzuela and Borja Mi\~nano for their assistance with the Simflowny platform. E.B., R.C and A.K. acknowledge support from the European Union's H2020 ERC Consolidator Grant ``GRavity from Astrophysical to Microscopic Scales'' (Grant No. GRAMS-815673), the PRIN 2022 grant ``GUVIRP - Gravity tests in the UltraViolet and InfraRed with Pulsar timing'', and the EU Horizon 2020 Research and Innovation Programme under the Marie Sklodowska-Curie Grant Agreement No. 101007855. Numerical calculations have been made possible through CINECA-INFN and SISSA-CINECA agreements providing access to resources on LEONARDO at CINECA.%(allocation INF24\_teongrav)
 MB acknowledge partial support from the STFC Consolidated
Grant no. ST/Z000424/1. This work used the DiRAC@Durham facility managed by the Institute for Computational Cosmology on behalf of the STFC DiRAC HPC Facility (www.dirac.ac.uk). The equipment was funded by BEIS capital funding via STFC capital grants ST/P002293/1, ST/R002371/1 and ST/S002502/1, Durham University and STFC operations grant ST/R000832/1. DiRAC is part of the National e-Infrastructure.

\appendix

\section{Solving for the radial function} \label{app:solve_mode_functions}

\begin{figure}[h]  
    \centering  % Center the image
    \includegraphics[width=0.5\textwidth]{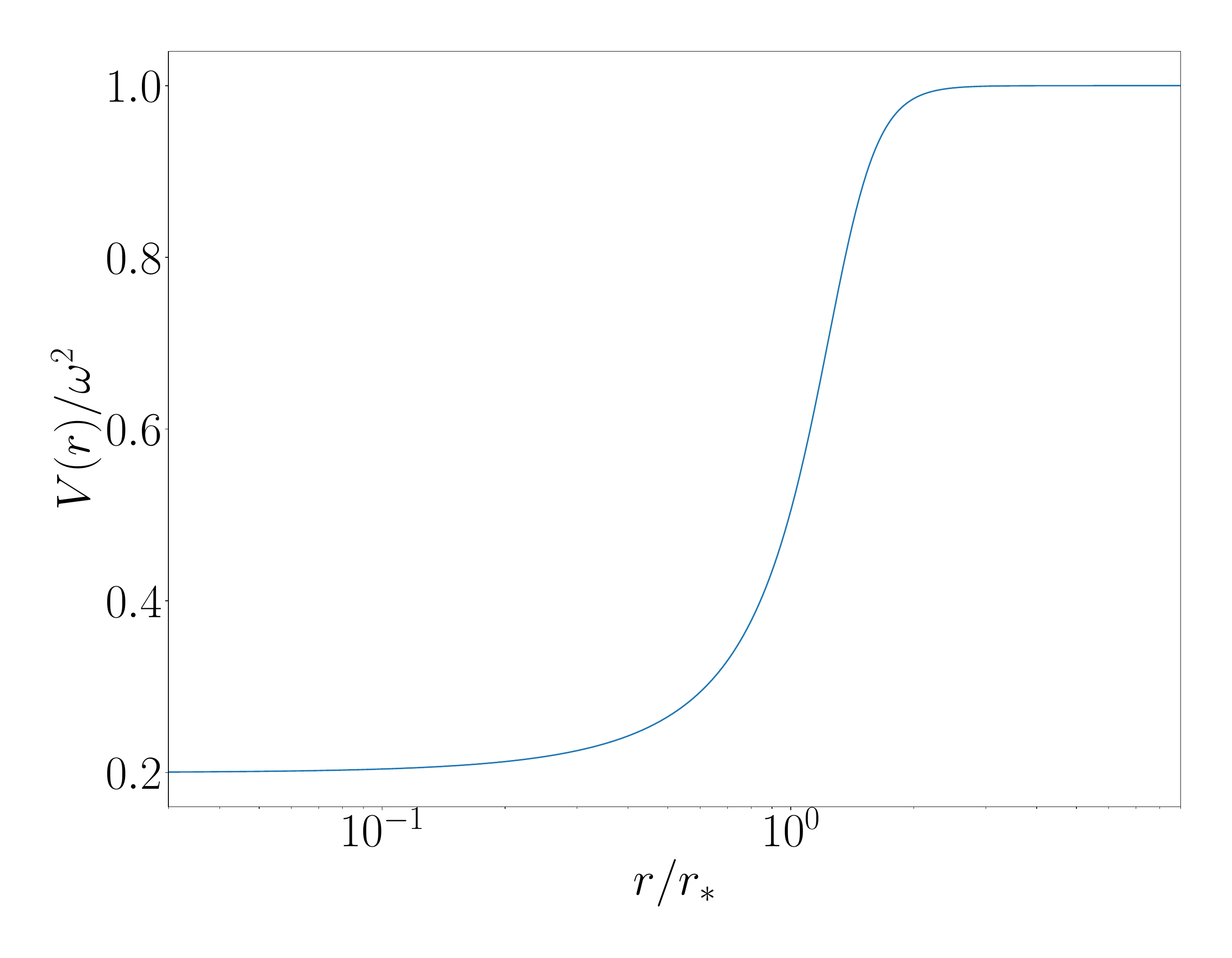} 
    \caption{ Plot of the potential $V(r)/\omega^2 = K'/(K'+2X_\mathrm{SS} K'')$ entering the equations of motion in the WKB regime, c.f. Eq.~\eqref{eq:EOMWKB}.   }  % Caption for the image
    \label{fig:potential}  % Label for referencing the image
\end{figure}

In this Appendix, we will construct the radial function $R$ by solving the equation of motion~\eqref{eq:EOMR} in two different regimes, which we then stitch together. The first regime in which one can analytically solve Eq.~\eqref{eq:EOMR} is the eikonal limit $\omega r \gg 1$. Note that we will still keep $r$ arbitrary with respect to the screening radius $r_*$, which is much larger than the wavelength of the radiation $\lambda = 2 \pi/\omega$. 
We change variables to define a rescaled radial function $\Psi$ as
\begin{equation}
    R(r) = \frac{g(r)}{r} \Psi(r) \; , \quad g(r) = \big( - K' - 2 X_\mathrm{SS} K'' \big)^{-1/2} \; .
\end{equation}
Inserting this change of variables into Eq.~\eqref{eq:EOMR}, we obtain
\begin{equation}
    \Psi'' + \bigg[ \frac{g''}{g} - 2 \bigg( \frac{g'}{g} \bigg)^2 + 2 \frac{g'}{rg} - K' g^2 \bigg( \omega^2 - \frac{\ell(\ell+1)}{r^2} \bigg) \bigg] \Psi = 0 \; .
\end{equation}
In the limit $\omega r \gg 1$,  most  terms inside the brackets are suppressed. We are thus left with a simple second-order equation on $\Psi$,
\begin{equation} \label{eq:EOMWKB}
    \Psi'' + \omega^2 \frac{K'}{K'+2X_\mathrm{SS}K''} \Psi = 0 \; .
\end{equation}
The potential $V(r) = \omega^2 K'/(K'+2X_\mathrm{SS} K'')$ is plotted in Fig.~\ref{fig:potential} for the  choice of $K(X)$ used in this work. It asymptotes to $\omega^2/5$ at distances $r \ll r_*$ smaller than the screening radius, and to $\omega^2$ outside  the screening radius ($r \gg r_*$). More importantly, $V(r)$ never vanishes: this means that we can use a Wentzel-Kramers-Brillouin (WKB) approximation to find a solution~\cite{Bender1978},
\begin{equation}
    \Psi_\mathrm{WKB}(r) = V(r)^{-1/4} \exp \bigg(+ i \int_0^r V(r')^{1/2} \mathrm{d}r'\bigg) \; ,
\end{equation}
where the plus sign has been chosen in the exponential in order for the solution to satisfy outgoing boundary conditions at infinity. 
This solution will be accurate as long as $V^{1/2} r \gg 1$. Thus, the solution for $R$ is
\begin{align}
    R(r) &= \frac{1}{r \omega^{1/2} \big( K'(K'+2X_\mathrm{SS} K'' )\big)^{1/4}} \nonumber \\
    &\times \exp \bigg( i \omega \int_0^r \sqrt{\frac{K'}{K'+2X_\mathrm{SS}K''}} \mathrm{d}r' \bigg) \; .
\end{align}
 In the following, it will be useful to have asymptotic expressions for $R$ at distances large and small relative to the screening radius. Using the solution for the spherically symmetric field in~\eqref{eq:phiSSInside}-\eqref{eq:phiSSOutside}, we obtain
\begin{align}
    R(r) &\simeq \frac{\sqrt{2} e^{i \omega r/\sqrt{5}}}{5^{1/4} \omega^{1/2} r^{1/5} r_*^{4/5}} \; , \quad r \ll r_* \; ,   \label{eq:R1WKBsmallr}%\\
     %R^{(2)}(r) &\simeq \frac{\sqrt{2} e^{-i \omega r/\sqrt{5}}}{5^{1/4} \omega^{1/2} r^{1/5} r_*^{4/5}}  \; , \quad r \ll r_* \label{eq:R2WKBsmallr}
\end{align}
where we have used the definition of the screening radius in~\eqref{eq:defrs}, and 
\begin{widetext}
\begin{align}
     R(r) &\simeq e^{i \omega r} \frac{\sqrt{2}}{r \omega^{1/2}} \exp \bigg( i \omega \int_0^\infty \bigg[ \sqrt{\frac{K'}{K'+2X_\mathrm{SS}K''}} - 1 \bigg] \mathrm{d}r' \bigg) \; , \quad  r \gg r_* \; . \label{eq:R1WKBlarger} %\\
    %R^{(2)}(r) &\simeq e^{-i \omega r} \frac{\sqrt{2}}{r \omega^{1/2}} \exp \bigg(- i \omega \int_0^\infty \bigg[ \sqrt{\frac{K'}{K'+2X_\mathrm{SS}K''}} - 1 \bigg] \mathrm{d}r' \bigg) \; , \quad  r \gg r_*
\end{align}
\end{widetext}
Note that the integral involved in the $r \gg r_*$ case is convergent, but cannot be computed analytically in closed-form. However, it is straightforward to numerically solve for the spherically symmetric field in Eq.~\eqref{eq:phiSS} and then compute the integral, obtaining
\begin{equation} \label{eq:approximateIntegral}
    \int_0^\infty \bigg[ \sqrt{\frac{K'}{K'+2X_\mathrm{SS}K''}} - 1 \bigg] \mathrm{d}r' \simeq - 0.56 r_* \; .
\end{equation}

The $\omega r \gg 1$ limit does not allow one to obtain the radial function in the wave zone. However,  there is another limit in which we can analytically compute the radial function, i.e. when the radius is small compared to the screening scale, $r \ll r_*$. In this case,  Eq.~\eqref{eq:EOMR} becomes
\begin{equation}
    r^{-2/5} \partial_r \big( r^{2/5} R' \big) + \frac{1}{5} \bigg( \omega^2 - \frac{\ell(\ell+1)}{r^2} \bigg) R = 0 \; .
\end{equation}
The outgoing solution is given by the Hankel function,
\begin{equation}
     R(r) = \mathcal{N} r^{3/10} H_\nu^{(1)} \bigg( \frac{\omega r}{\sqrt{5}} \bigg) \; ,
\end{equation}
where $\mathcal{N}$ is a normalization constant, and $\nu = \sqrt{9+20\ell(\ell+1)}/10$. We fix $\mathcal{N}$ by requiring that the solution should be equal to the WKB solution in the matching zone $1/\omega \ll r \ll r_*$. We will thus need the asymptotic expansion of $H_\nu^{(1)}$ for $\omega r \gg 1$, given by
\begin{align}
    R(r) &\simeq - \mathcal{N} \frac{e^{i \omega r/\sqrt{5}}}{\omega^{1/2} r^{1/5}} \sqrt{\frac{2 \sqrt{5}}{\pi}} e^{i \pi (3/4-\nu/2)} \; , \quad r\omega \gg 1 \; . %\\
    %R^{(2)}(r) &\simeq - i \mathcal{N}^{(2)} \frac{e^{-i \omega r/\sqrt{5}}}{\omega^{1/2} r^{1/5}} \sqrt{\frac{2 \sqrt{5}}{\pi}} e^{i \pi (3/4+\nu/2)} \; , \quad r\omega \gg 1
\end{align}
Comparing with Eq.~\eqref{eq:R1WKBsmallr}, we obtain the value of the normalization factor,
\begin{align}
    \mathcal{N} &= - \sqrt{\frac{\pi}{5}} e^{i \pi (\nu/2-3/4)}  r_*^{-4/5} \; . \label{eq:N1} %\\
    %\mathcal{N}^{(2)} &= i \sqrt{\frac{\pi}{5}} e^{-i \pi (\nu/2+3/4)} r_*^{-4/5}
\end{align}
To summarize, we have obtained the radial function for all values of the radius:
\begin{align}
    R &= \frac{1}{r \omega^{1/2} \big( K'(K'+2X_\mathrm{SS} K'' )\big)^{1/4}} \nonumber \\ 
    &\times \exp \bigg( i \omega \int_0^r \sqrt{\frac{K'}{K'+2X_\mathrm{SS}K''}} \mathrm{d}r' \bigg) \; , \; r \omega \gg 1 \; , \label{eq:R1WKB} \\
    R &= \mathcal{N} r^{3/10} H_\nu^{(1)} \bigg( \frac{\omega r}{\sqrt{5}} \bigg) \; , \quad r \ll r_* \; . \label{eq:R1smallr}
\end{align}
%and similarly for $R^{(2)}$.

\section{Matching multipole and near-zone solutions} \label{app:matching}

In this Appendix, we explicitly perform the matching of the near-zone solution obtained in Sec.~\ref{sec:static1body}, Eq.~\eqref{eq:PNsol} (valid for $|\bm x | \ll 2 \pi/\Omega$), and the multipole expansion obtained in Sec.~\ref{sec:multipole}, $\varphi = \varphi_\mathrm{SS} + \varphi_1$, where $\varphi_1$ is given by Eq.~\eqref{eq:multipole_decomposition}. 
%The field at large distances $\varphi = \varphi_\mathrm{SS} + \varphi_1$, where $\varphi_1$ is given in Eq.~\eqref{eq:phiLarger}, is given as a multipole expansion over the angles $(\theta, \phi)$. 
To this aim, let us perform a multipole expansion of the near-zone solution~\eqref{eq:PNsol}, and match the coefficients of these decompositions. 

To perform the multipole expansion, we note that at large distances from the source $|\bm x| \gg a$ (but still inside the near zone $|\bm x| \ll 2 \pi/ \Omega$) we can write, at lowest order,
\begin{align}
    \varphi(\bm x) &= \varphi_\mathrm{SS}( | \bm x - \bm x_\mathrm{NS}(t) |) \nonumber \\ 
    &\simeq \varphi_\mathrm{SS}( | \bm x |) - x_\mathrm{NS}^i n_i \varphi_\mathrm{SS}'( | \bm x |) + \dots \label{eq:multipole_expansion_wave_zone}
\end{align}
where $n_i = x_i/|\bm x|$. The only angular dependence is in the scalar product $x_\mathrm{NS}^i n_i$, whose decomposition in spherical harmonics gives a dipole term:
\begin{align}
    &x_\mathrm{NS}^i n_i = - \frac{m_\mathrm{BH}}{m_\mathrm{BH}+m_\mathrm{NS}} a \sin \theta \cos (\Omega t - \phi) \\
    &= \sqrt{\frac{2 \pi}{3}} \frac{m_\mathrm{BH}}{m_\mathrm{BH}+m_\mathrm{NS}}  a  Y_{11}(\theta, \phi) e^{-i \Omega t}  + \mathrm{c.c.}
\end{align}
where $\theta$ and $\phi$ are the angles of the observation direction $n_i$ in spherical coordinates.
Let us now perform the matching. The monopole term $\varphi_\mathrm{SS}(| \bm x|)$ is the same in the near-zone solution [Eq.~\eqref{eq:multipole_expansion_wave_zone}] and in the multipole expansion [Eq.~\eqref{eq:multipolar_expansion}] provided that we set $\mathcal{A}_{00\omega} = 0$. 
On the other hand, the dipole term will match if
\begin{align} \label{eq:matching}
    \int \mathrm{d}\omega e^{-i \omega t} \mathcal{A}_{11 \omega} R_{11 \omega}(r) &= - \sqrt{\frac{2 \pi}{3}}  \frac{m_\mathrm{BH}}{m_\mathrm{BH}+m_\mathrm{NS}} \nonumber \\
    &\times a e^{-i \Omega t}  \varphi_\mathrm{SS}'( r) \; .
\end{align}
At this order in the expansion in $a/r$, we can set all other multipoles to zero. 

Let us now see how we can satisfy this matching condition. As before, we assume that the near zone is completely contained inside the screening radius $r_*$, i.e. $2 \pi/ \Omega \ll r_*$, which is very well verified for typical astrophysical sources. The radial function $R$ for $r \ll r_*$ has been determined in App.~\ref{app:solve_mode_functions}, c.f. Eq.~\eqref{eq:R1smallr}. We can then further simplify the latter, because in the wave zone $\omega r \ll 1$:
\begin{equation}
    R_{11\omega}(r) \simeq \frac{- i \mathcal{N}}{\pi r^{2/5}} \Gamma(7/10) \bigg( \frac{2 \sqrt{5}}{\omega} \bigg)^{7/10} \; ,
\end{equation}
where we have kept the most diverging term in the Hankel function, and $\mathcal{N}$ is given by Eq.~\eqref{eq:N1}. On the other hand, in the screening region we have [from Eq.~\eqref{eq:phiSSInside}]
\begin{equation}
    \varphi'_\mathrm{SS} = \left(  \frac{\alpha_\mathrm{NS} m_\mathrm{NS} \Lambda^8 }{3 \pi \mpl r^2 } \right)^{1/5} \; .
\end{equation}
Thus, the $r$-dependence of the multipole solution and of the near-zone solution in Eq.~\eqref{eq:matching} is the same. 

Therefore, one just needs to match the constant $ \mathcal{A}_{11 \omega}$. Let us factor out a delta function ensuring the matching of the time-dependent terms in Eq.~\eqref{eq:matching}, $\mathcal{A}_{11\omega} =  \delta(\omega - \Omega) \tilde{\mathcal{A}}_{1 1 \Omega}$. We find
\begin{align}
     \tilde{\mathcal{A}}_{11 \Omega} &= i  e^{ 2i \pi/5} \sqrt{\frac{10}{3}} \frac{\alpha_\mathrm{NS} m_\mathrm{NS}}{4 \Gamma(7/10) \mpl} \nonumber \\
     &\times r_*^{-4/5} \bigg( \frac{\Omega}{2 \sqrt{5}} \bigg)^{7/10}  \frac{m_\mathrm{BH}}{m_\mathrm{BH}+m_\mathrm{NS}}  a \; ,
\end{align}
which is the expression presented in Sec.~\ref{sec:multipole}.
%The constant $ \mathcal{A}_{1-1 \omega}$ takes the same expression, with the replacement $\delta(\omega - \Omega) \rightarrow - \delta(\omega + \Omega)$. 

\section{Solution for the scalar in different regimes} \label{app:different_regimes}

In this Appendix, we will study the behavior of the dipole scalar~\eqref{eq:phi1} in three different asymptotic zones, in order to explain the scaling observed in Fig.~\ref{fig:dipole}: in the wave zone ($a \ll r \ll 2 \pi/\Omega$); between the wave zone and the screening radius ($2 \pi/\Omega \ll r \ll r_*$); and outside the screening radius ($r \gg r_*$). 
Let us discuss separately these three cases. We will pay  particular attention to the scaling of the field in each of these regimes, and check that our basic assumption $\varphi_1 \ll \varphi_\mathrm{SS}$ is always valid as long as $r \gg a$, thus proving the existence of a buffer zone \textit{a posteriori}. 

\subsection{$a \ll r \ll 2 \pi/\Omega$}

In this case, we have already determined the field in Eq.~\eqref{eq:matching}, so we have
\begin{equation}
     \tilde{\mathcal{A}}_{11 \Omega}  R_{1 1 \Omega}(r) \simeq - \sqrt{\frac{2 \pi}{3}} a \frac{m_\mathrm{BH}}{m_\mathrm{NS}+m_\mathrm{BH}}  \bigg( \frac{\alpha_\mathrm{NS} m_\mathrm{NS} \Lambda^8}{3 \pi \mpl r^2} \bigg) ^{1/5} \; .
\end{equation}
Let us do a simple order-of-magnitude estimate, neglecting all order-one factors and setting $m_\mathrm{BH}/(m_\mathrm{NS}+m_\mathrm{BH}) \simeq 1$. Normalizing the field to the background value of the scalar $\varphi_\mathrm{SS}$, we obtain
\begin{equation}
    \frac{\varphi_1}{\varphi_\mathrm{SS}} \sim \frac{a}{r} \ll 1 \; . \label{eq:order-of-mag-r<<omega}
\end{equation}
This is the standard result for a dipole field in the wave zone, and  shows that our assumption $\varphi_1 \ll \varphi_\mathrm{SS}$ is justified in this regime.

\subsection{$2 \pi/\Omega \ll r \ll r_*$}

In this case, we can use Eq.~\eqref{eq:R1WKBsmallr} to obtain
\begin{align}
     &\tilde{\mathcal{A}}_{11 \Omega}  R_{1 1 \Omega}(r) \simeq \frac{ i  e^{2 i \pi/5} }{\Gamma(7/10) 5^{7/10} 2^{11/5} 3^{1/2}} \frac{e^{i \Omega r/\sqrt{5}}}{r^{1/5} r_*^{8/5}} \nonumber \\
     &\times \frac{\alpha_\mathrm{NS} m_\mathrm{NS}}{\mpl} a \frac{m_\mathrm{BH}}{m_\mathrm{NS}+m_\mathrm{BH}} \Omega^{1/5} \; .
\end{align}
The order-of-magnitude estimate gives
\begin{equation}
    \frac{\varphi_1}{\varphi_\mathrm{SS}} \sim a \bigg( \frac{\Omega}{r^4} \bigg)^{1/5} \sim v^{1/5} \bigg( \frac{a}{r} \bigg)^{4/5} \ll 1 \; ,\label{eq:order-of-mag-omega<<r<<rs}
\end{equation}
where we have introduced the typical velocity of the point particle $v=\Omega a$. Note that the two order-of-magnitude estimates in Eqs.~\eqref{eq:order-of-mag-r<<omega} and~\eqref{eq:order-of-mag-omega<<r<<rs} give the same value at the boundary of their respective domains, $r \sim 2 \pi/\Omega$: $\varphi_1/\varphi_\mathrm{SS} \sim \Omega a$. Note also that the estimate~\eqref{eq:order-of-mag-omega<<r<<rs} implies that the dipole term  continues to fall off with $r$ more quickly than the background field $\varphi_\mathrm{SS}$ inside the screening radius, while the ratio $\varphi_1/\varphi_\mathrm{SS}$ would be a (small) constant for a FJBD field outside the near zone. This indicates that the field is probably screened outside the screening radius, as we will now confirm.

\subsection{$r \gg r_*$}

In this case, we use Eq.~\eqref{eq:R1WKBlarger} to obtain
\begin{align}\label{eq:dipole_pred}
    &\tilde{\mathcal{A}}_{11 \Omega}  R_{1 1 \Omega}(r) \simeq i  e^{ 2i \pi/5} \frac{5^{3/20}}{\Gamma(7/10) 2^{17/10} 3^{1/2}} \frac{e^{i \Omega (r-0.56 r_*)}}{r} \nonumber   \\
    &\times  \frac{\alpha_\mathrm{NS} m_\mathrm{NS}}{\mpl} \frac{\Omega^{1/5}}{r_*^{4/5}}  a \frac{m_\mathrm{BH}}{m_\mathrm{BH}+m_\mathrm{NS}}  \; , 
\end{align}
where the factor of $0.56$ comes from the numerical evaluation of the integral in Eq.~\eqref{eq:approximateIntegral}. 
An order-of-magnitude estimate gives
\begin{equation} \label{eq:order-of-mag-r>>rs}
     \frac{\varphi_1}{\varphi_\mathrm{SS}} \sim v^{1/5} \bigg( \frac{a}{r_*} \bigg)^{4/5} \ll 1 \; ,
\end{equation}
where we have used the definition of the screening radius in Eq.~\eqref{eq:defrs}. Note once again that the two order-of-magnitude estimates~\eqref{eq:order-of-mag-omega<<r<<rs} and~\eqref{eq:order-of-mag-r>>rs} agree at their common boundary $r \sim r_*$.

\section{Matter profile dependence} \label{app:mattdep}

\begin{figure}[ht]  
    \centering  % Center the image
    \includegraphics[width=0.475\textwidth]{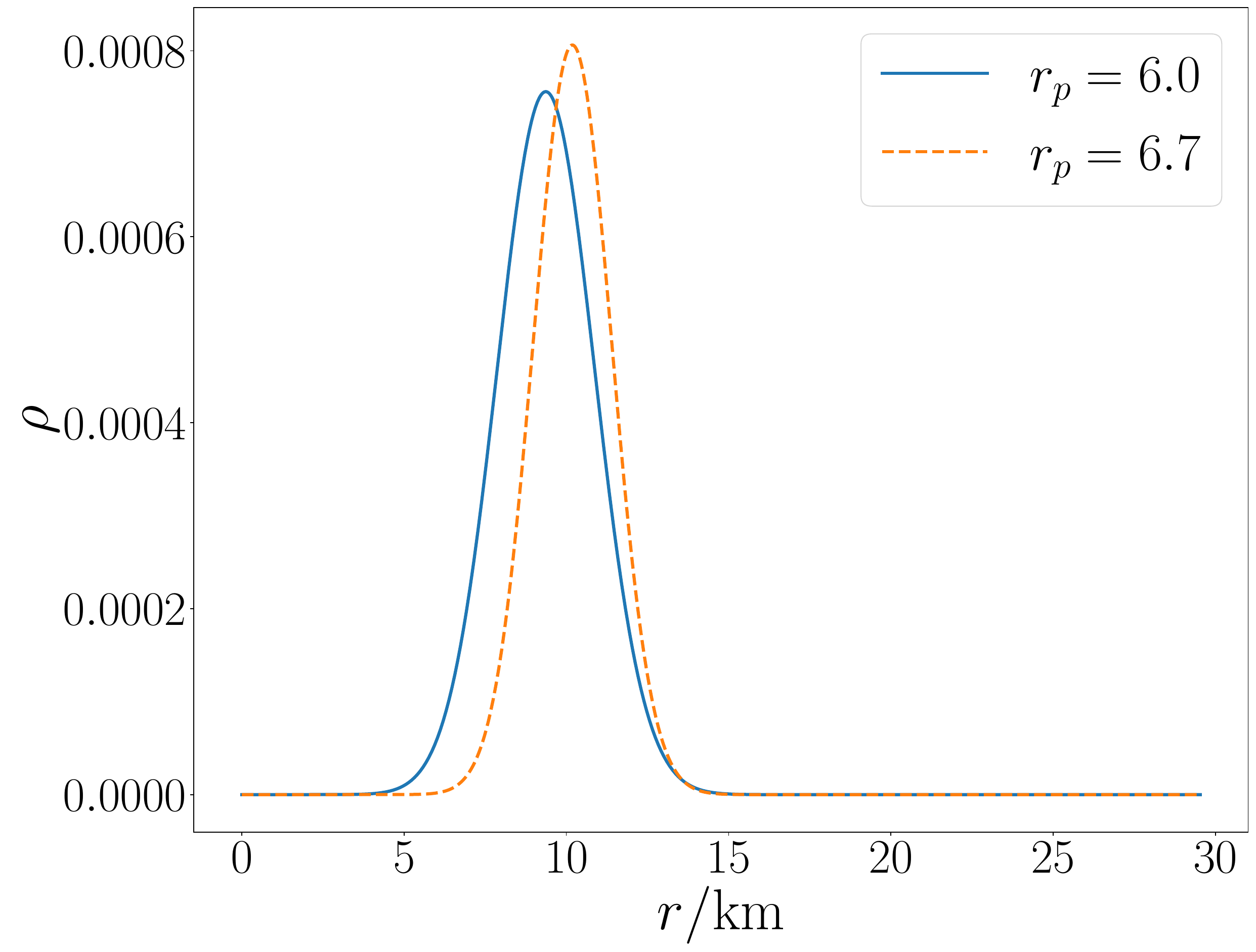}  % Adjust the width as needed
    \caption{Density profile $\rho$ for the neutron star as a function of the distance from the center  $r$, for two choices of the $r_{p}$ parameter.}   % Caption for the image
    \label{fig:shells}  % Label for referencing the image
\end{figure}

In this Appendix, we show that the outgoing radiation is insensitive to the profile chosen to describe the matter source $\rho$. To check this, we take the definition \eqref{eq:matprofile} and choose two different values of $r_p$. Given the chosen value of $r_{p}$, we adjust $\sigma$ so that $99\%$ of the mass is contained in the radius $r_s=13.3 \mathrm{km}$, and also adjust $A$ so that the total mass of the star is $m_{NS}=1M_{\odot}$. In Fig.~\ref{fig:shells}, we show two such choices that produce considerably different profiles for $\rho$. 

\begin{figure}[ht]  
    \centering  % Center the image
    \includegraphics[width=0.475\textwidth]{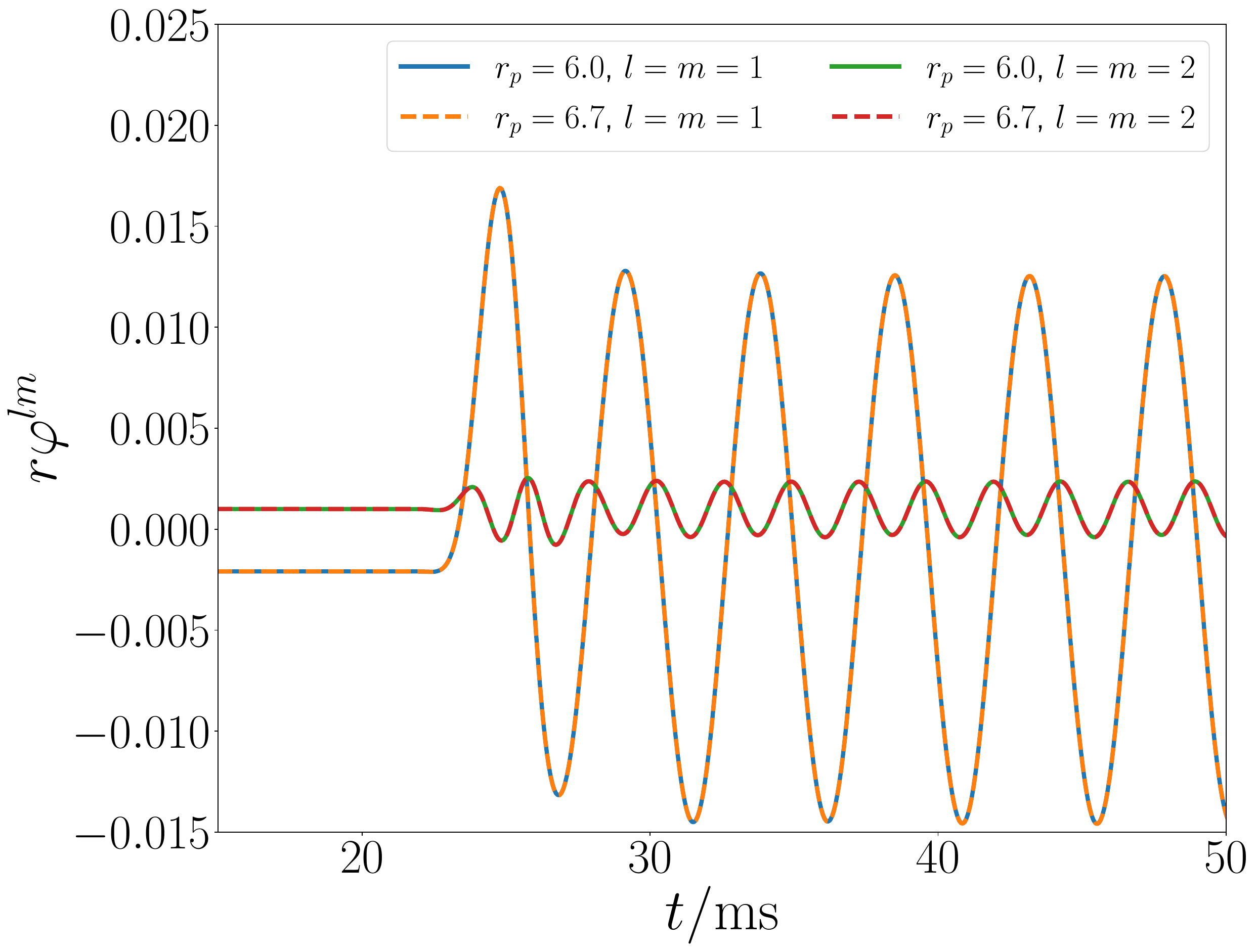}  % Adjust the width as needed
    \caption{Waveforms for the dipole $l=m=1$  and quadrupole $l=m=2$ of the outgoing scalar radiation, extracted at a radius of $r=7383 \mathrm{km}$ and for different choices of the matter profile $\rho$.}   % Caption for the image
    \label{fig:shells_dip_quad}  % Label for referencing the image
\end{figure}

\begin{figure}[ht]  
    \centering  % Center the image
    \includegraphics[width=0.475\textwidth]{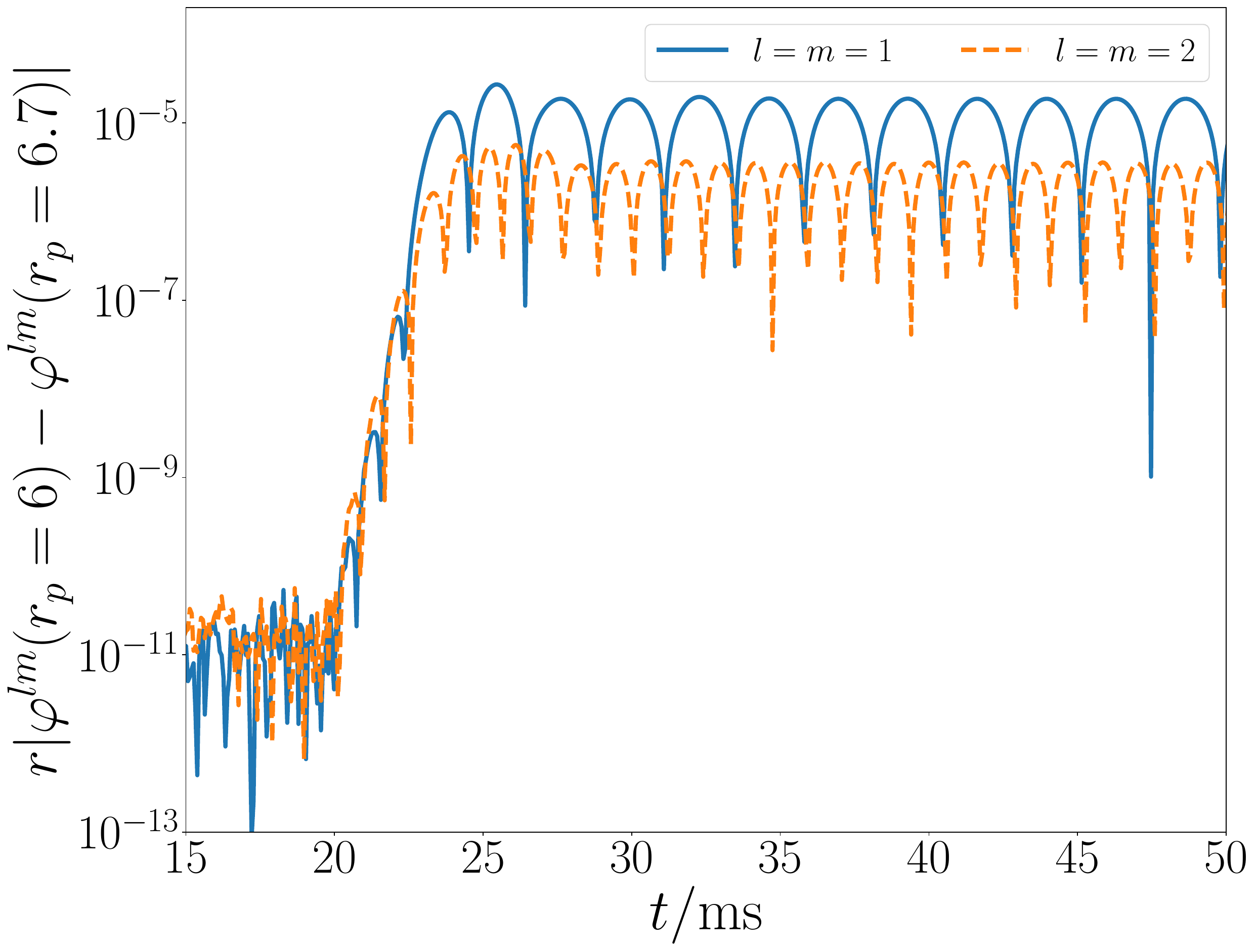}  % Adjust the width as needed
    \caption{Absolute difference between waveforms for the dipole $l=m=1$  and quadrupole $l=m=2$ of the outgoing scalar radiation,  extracted at a radius of $r=7383 \mathrm{km}$ and for different choices of the matter profile $\rho$.}   % Caption for the image
    \label{fig:errorshells_dip_quad}  % Label for referencing the image
\end{figure}

Then, for these two choices of the profile, we carry out numerical evolutions (for $\Lambda=0.69 \mathrm{MeV}$). We compare the outgoing dipole and quadrupole radiation at a fixed radius, as shown in Fig.~\ref{fig:shells_dip_quad}, and the absolute difference between the two in Fig.~\ref{fig:errorshells_dip_quad}. These plots confirm that the outgoing radiation is mainly independent of the source profile, supporting the robustness of our results.

\section{Convergence} \label{app:convergence}
To test the convergence of our results, we compare our finest-grid solution ($\Delta x_{\rm{low}} \equiv \Delta x_{10} = 0.123 \mathrm{km}$) with two other higher resolutions  $\Delta x _{\rm{med}} = 0.0984 \mathrm{km}$ and $\Delta x _{\rm{high}} = 0.082 \mathrm{km}$. The value of the strong coupling scale used is $\Lambda=0.69\mathrm{MeV}$. The error between two resolutions is calculated on the quadrupole  $l=m=2$ waveforms as a function of time and is defined as $E^{h1-h2}\equiv |r\varphi^{l=m=2}_{\Delta x_{h1}} - r\varphi^{l=m=2}_{\Delta x_{h2}}|$. Fig.~\ref{app:convergence} shows the error between the low and medium resolutions $E^{\rm{low}-\rm{med}}$ and between the medium and high resolution $E^{\rm{med}-\rm{high}}$. This figure also displays the estimates for third and fourth-order convergence for $E^{\rm{low}-\rm{med}}$ based on the proportionality factor
\begin{equation}
\mathcal{Q}_{n} = \frac{\Delta x_{\rm{low}}^{n}-\Delta x_{\rm{med}}^{n}}{\Delta x_{\rm{med}}^{n}-\Delta x_{\rm{high}}^{n}},    
\end{equation}
where $n$ is the order of convergence. The estimate for the error between low resolution and medium resolution assuming order of convergence $n$ is given by $E_{n}^{\rm{low}-\rm{med}}\equiv \mathcal{Q}_{n} E^{\rm{med}-\rm{high}}$. In this figure, we can see how the $E^{\rm{low}-\rm{med}}$ falls in between the predictions for third and fourth-order convergence $E_{3}^{\rm{low}-\rm{med}}$ and $E_{4}^{\rm{low}-\rm{med}}$ respectively. This indicates approximately third-order convergence.
\begin{figure}[ht]  
    \centering  % Center the image
    \includegraphics[width=0.475\textwidth]{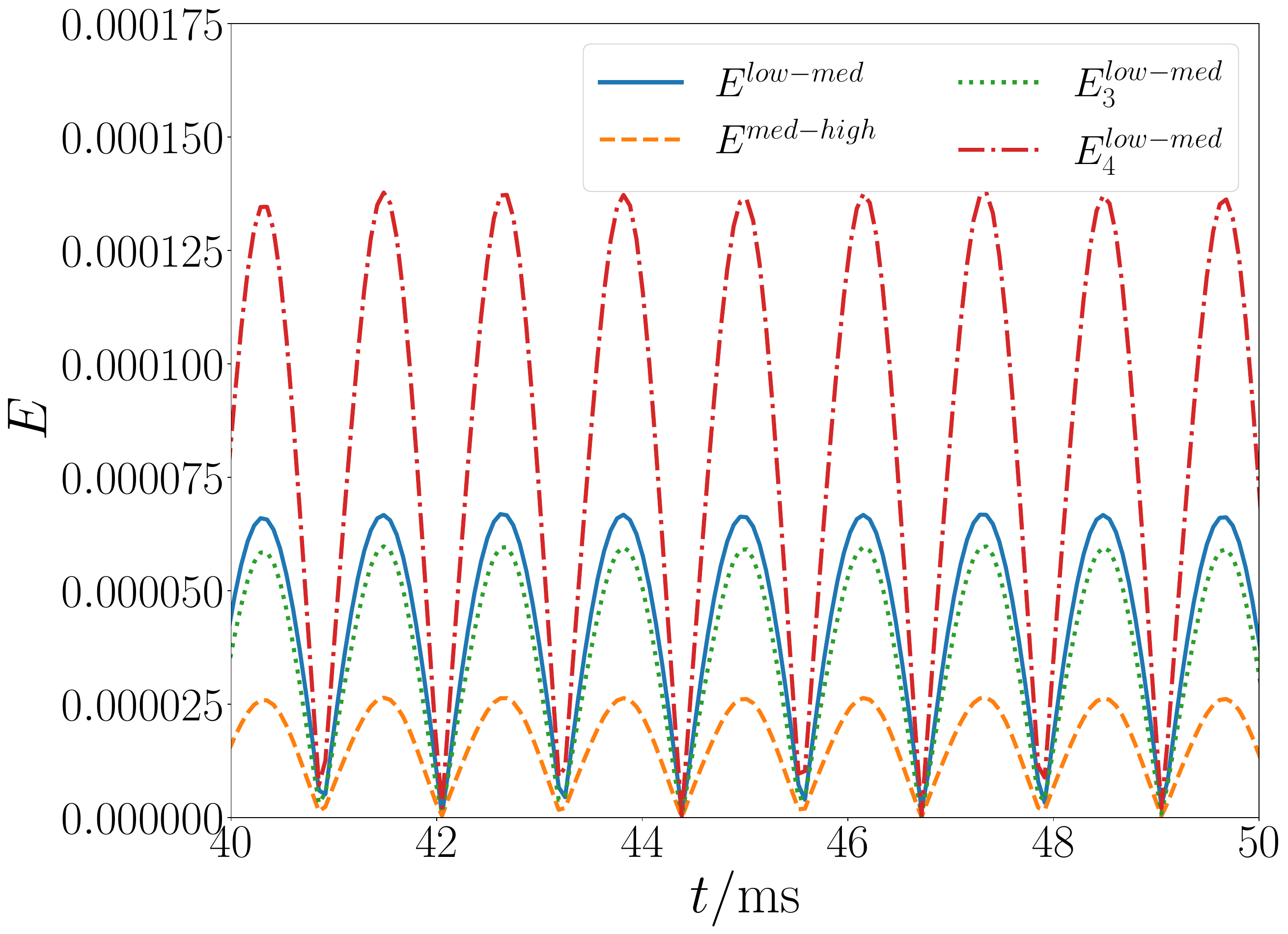}  % Adjust the width as needed
    \caption{Errors for the quadrupole $l=m=2$ waveforms extracted extracted at a radius of $r=7383 \mathrm{km}$ for $\Lambda=0.69\mathrm{MeV}$. The green dotted and red dashed-dotted lines show estimates $E_{3}^{\rm{low}-\rm{med}}$ and $E_{4}^{\rm{low}-\rm{med}}$ of the error between the low and medium resolution, assuming third and fourth-order convergence, respectively. These curves are consistent with approximately third-order convergence.}   % Caption for the image
    \label{fig:convergence}  % Label for referencing the image
\end{figure}

\bibliography{bib}

\end{document}